\documentclass[authoryear,3p,12pt]{elsarticle}

\usepackage[utf8]{inputenc}

\usepackage{natbib}

\usepackage{amsthm,amsmath}

\usepackage{graphicx}
\usepackage{subfig}

\usepackage{nameref}
\usepackage{lmodern}

\usepackage[]{algorithm2e}

\newcommand{\Chi}{\mathcal{X}}

\journal{arXiv.org}
\begin{document}
\begin{frontmatter}
\title{THERMAID - A matlab package for thermo-hydraulic modeling and fracture stability analysis in fractured reservoirs}
\author[unine]{Gunnar Jansen\corref{cor1}}\fnref{fn1}
\ead{gunnar.jansen@unine.ch}
\author[unine]{Benoît Valley}\fnref{fn1}
\ead{benoit.valley@unine.ch}
\author[unine]{Stephen A. Miller}\fnref{fn1}
\ead{stephen.miller@unine.ch}

\cortext[cor1]{Corresponding author}
\address[unine]{CHYN - Center for Hydrogeology and Geothermics, Laboratory of Geothermics and Geodynamics, University of Neuchâtel, Switzerland}

\fntext[fn1]{Author's contributions: G.J. and B.V. designed the model and the	computational framework. G.J. carried out the implementation and performed the calculations. G.J	wrote the manuscript in consultation with B.V. and S.A.M. S.A.M. conceived the study and was in charge of overall direction and planning.}

\begin{abstract}
Understanding the dynamics of naturally fractured systems and fractured reservoirs in terms of flow, heat transport and fracture stability (e.g. induced seismicity) is important for a range of applications associated with waste water injection, renewable energy (e.g. geothermal systems), and greenhouse gas mitigation (e.g. geological sequestration of CO2). Here we present the implementation and validation of an open source MATLAB package for efficient numerical simulations of the coupled processes in fractured systems. We take advantage of the embedded discrete fracture model that efficiently accounts for discrete fractures. We perform a series of numerical benchmark experiments to validate the implemented approach against analytical solutions and established numerical methods. Finally, we use a simplified geomechanical model and an integrated fracture stability analysis that allows estimating the potential for shear stimulation, and thus a mechanistic assessment of induced seismic risk  during stimulation. The open source distribution of the source code and results can be used as a blue print for the re-implementation of the method in a high performance computing (HPC) framework or as a standalone simulation package for investigating TH(m) problems in geothermal reservoirs.
\end{abstract}


\end{frontmatter}

\section{Introduction}
A large fraction of the world's water and energy resources are located in naturally fractured reservoirs within the earth's crust. Understanding the dynamics of such reservoirs in terms of flow, heat transport and fracture stability is crucial to successful application of engineered geothermal systems (also known as enhanced geothermal systems, EGS) for geothermal energy production. Reservoir development characteristics such as permeability creation and induced seismicity largely depend on the properties of preexisting fractures, porosity, permeability and fracture orientation within the local stress field. One of the primary driving mechanisms for permeability creation in EGS involves shear failure induced by fluid injection at high pressures \citep{barton1995,evans2005,hickman2010}. Along sections of the well that are free of natural fractures and in environments with low differential stress, tensile fractures may develop if the injection pressure exceeds the minimal principal stress. Shear and tensile fracture propagation and reactivation are not exclusive and might occur simultaneously during the stimulation of the reservoir \citep{mcclure2014}. Clearly, preexisting, critically stressed and optimally oriented fractures provide the most favorable conditions for enhancing permeability of EGS  \citep{barton1995, combs2004, ghassemi2007}.\\
The basis for EGS are fractured reservoirs, which are usually geothermal plays of the ``hot dry rock'' type where the available water in the porous medium is considered negligible \citep{brown2012}. These conditions are found primarily in metamorphic or igneous terrains with low permeability and porosity, containing fractures and faults that provide the major pathways for fluid flow (e.g. Fenton Hill, Soultz, Basel, Cooper basin and Desert Peak \citep{kelkar2016,hooijkaas2006,haring2008,chen2009, hickman2010}). In geothermal energy systems, the fracture's surfaces serve as the main heat exchanger. Fractured reservoirs can be considered to consist of two distinct separate media, the fractures and the matrix, and different types of reservoirs can be defined that depend on their properties  \citep{nelson2001}. In EGS, two cases typically prevail: 1) reservoirs with low porosity matrix for which both the permeability and the storage capacity of the rock mass are controlled by the fractures (cf. type 1 in \citep{nelson2001}) and 2) reservoirs with sufficient matrix porosity such that fluid storage is dominated by the matrix while the fractures contain only a small fraction of the fluid but control the permeability (cf. type 2 in \citep{nelson2001}). 

Simulation of flow and transport through fractured porous media is challenging due to the high permeability contrast between the fractures and the surrounding rock matrix. However, accurate and efficient simulation of flow through a fracture network is crucial in order to understand, optimize and engineer reservoirs. Even after decades of research, this is still a very active research topic. Additionally, accurate estimations of the fracture stability are necessary in order to predict permeability evolution and forecast induced seismicity. Discrete fracture models (DFM) have been developed to address the computational problem of scales for fluid flow and heat transport. Various modeling frameworks for the simulation of fractured porous media in general and geothermal reservoirs in particular exist. Literature reviews of current modeling approaches in geothermal reservoirs and hot dry rock systems are presented by Willis-Richards and Wallroth, Sanyal et al. and O’Sullivan et al.\citep{willis1995,sanyal2000, osullivan2001}. Some better known open-source modeling frameworks related to this work include PFLOTRAN \citep{pflotran-user-ref}, OpenGeoSys \citep{kolditz2012} and DuMux \citep{flemisch2011}. Yet traditional conforming DFM, where the fractures are explicitly resolved by the numerical grid, suffer from computationally expensive pre-processing in the numerical grid generation and can encounter severe time step restrictions during the simulation when using explicit time-stepping and small cells around the fractures  \citep{norbeck2014,sandve2012}. \\
An alternative approach uses the embedded discrete fracture models (EDFM), which treat fracture and matrix in two separate computational domains. The embedded fracture model was first introduced by Lee et al. for single phase problems and later extended to two-phase flow \citep{lee2001, li2008}. The embedded discrete fracture model is a promising technique in modeling the behavior of enhanced geothermal systems. Karvounis \citep{karvounis2013} employs EDFM and a statistical approach to better understand and possibly forecast seismicity induced seismicity by fluid injection during the stimulation phase of an EGS. Norbeck et al. additionally model fracture deformation by linear fracture mechanics \citep{norbeck2016}. 

In this paper we present the to our best knowledge first open source implementation of an embedded discrete fracture model for single phase flow and heat transport with additional capabilities to determine fracture stability in fractured reservoirs. Slip tendency analysis is used in order to estimate fault reactivation potential in earthquake prone areas as well as fracture stability in geothermal reservoirs \citep[e.g.][]{morris1996, moeck2009}. Slip tendency is an indicator for the likelihood of slip. Using slip tendency, predictions on fracture instabilities during the hydraulic stimulation of a fractured reservoir are feasible without solving for the typically non-linear evolution of the stress equilibrium equation. THERMAID, an acronym for "Thermo-Hydraulic Energy Resource Modeling for Application and Development", is a fractured reservoir modeling framework implemented in \textit{MATLAB}, which can be used as a standalone simulation package for TH(m) cases in geothermal reservoirs or as a blue print for the re-implementation of the method e.g. in a high performance computing (HPC) framework. We coin the term TH(m) to indicate a coupled Thermo-Hydraulic code, and we use the lower case (m) to indicate simplified mechanics.\\
This paper is structured as follows. In the next section we present the methodology of the embedded discrete fracture model, and describe in detail the underlying theory of the fracture stability analysis. The implemented model is evaluated in the \nameref{sec1:results} section by comparing it with a widely used numerical model in several test cases. We conclude the paper by illustrating possible applications of the code using some examples and a discussion of the findings.

\section{Methodology}\label{sec1:method}
\begin{figure}[!htbp]
	\centering
	\includegraphics[width=0.55\linewidth]{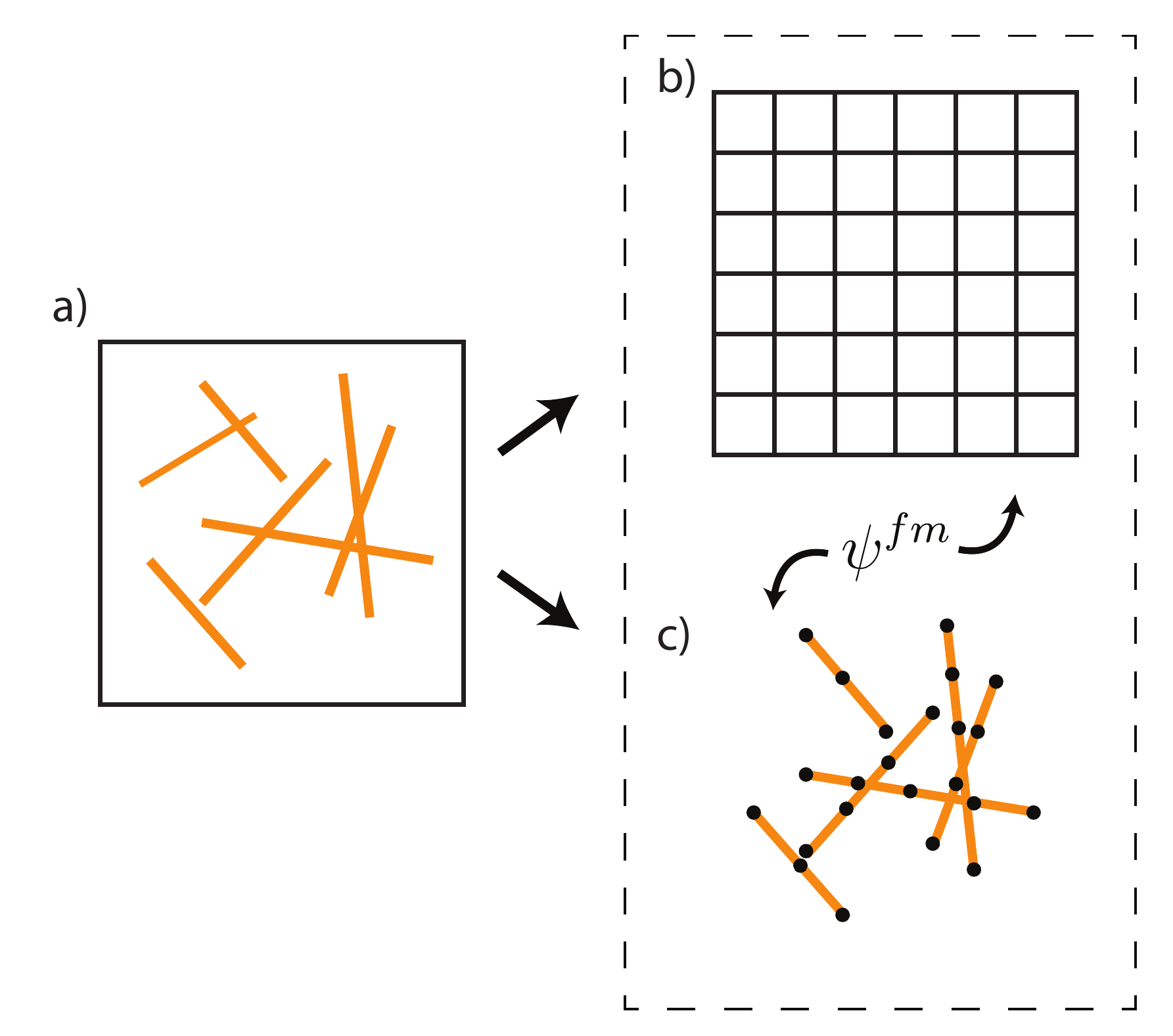}
	\caption{A fractured domain a) is separated in a uniform grid b) and a fracture grid c). The two resulting domains are coupled using the transfer function $\Psi^{fm}$ .}
	\label{fig1:edfm_domain}
\end{figure}

The conceptual idea of the EDFM is the distinct separation of a fractured reservoir into a fracture and a matrix domain. We introduce a transfer function to account for coupling effects between the two domains (cf. Figure \ref{fig1:edfm_domain}), so  the fracture and matrix domains are computationally independent except for the transfer function. As the fractures are generally very thin and highly permeable compared to the surrounding matrix rock, the gradient of fluid pressure with the fracture normal to it is negligible. This allows for a lower dimensional representation of fractures (i.e. 1D objects within a 2D reservoir).

\subsection{Conceptual model}
Numerical modeling of fractured reservoirs is not only challenging from the numerical and computational point of view, but also because it involves a variety of coupled thermal, hydraulic, mechanical and chemical (THMC) processes. THERMAID focuses on thermo-hydraulic processes and their coupling, with some additional mechanical processes considered within a simplified geomechanical model.  

\begin{figure}[!b]
	\centering
	\includegraphics[width=0.99\linewidth]{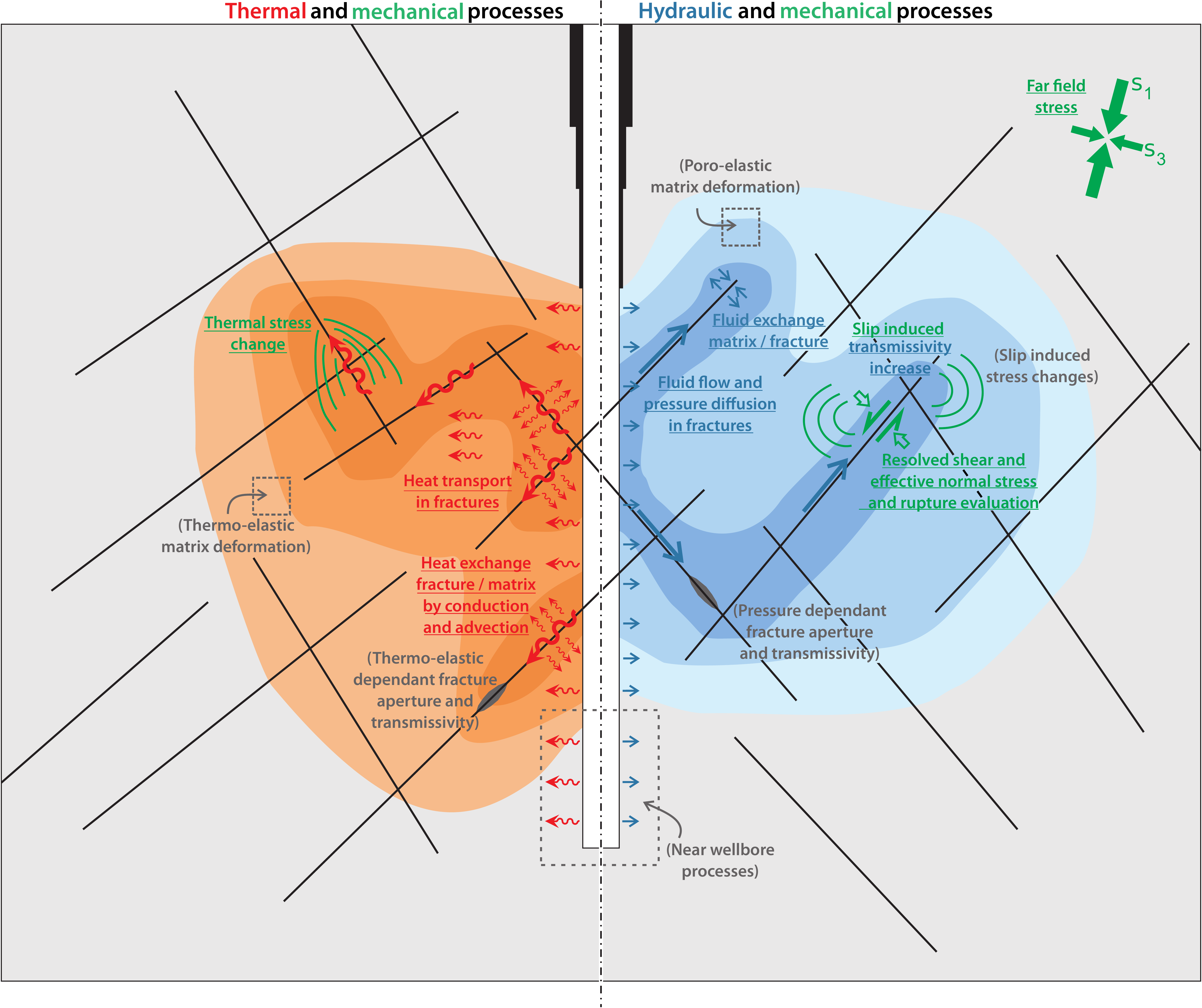}
	\caption{A conceptual model of a fractured domain with relevant thermo-hydro-mechanical processes. Underlined processes are included in THERMAID's current implementation. Processes in brackets are considered relevant but are currently not included mainly due to ambiguity in the simplified geomechanical model.}
	\label{fig1:concept}
\end{figure}
Figure \ref{fig1:concept} shows the conceptual model of the most relevant thermal, hydraulic and mechanical processes in fractured reservoirs. The core processes implemented in THERMAID are fluid flow through pressure diffusion in the matrix and the fracture network and the accompanying heat transfer by advection and diffusion. Pressure and heat are also exchanged between the rock matrix and the fracture at the fracture walls by pressure and thermal diffusion and thermal advection, respectively.\\
Associated with the thermo-hydraulic processes, numerous thermo-mechanic or thermo-hydraulic processes could be activated, but only a limited amount and simplified processes are  currently implemented in THERMAID. Poro-elastic and thermo-elastic deformation of the fractures and of the matrix is currently not implemented in the code. The fluid pressure in the fracture is, however, considered in the computation of fracture effective normal stress, which is an important parameter to evaluate fracture stability. If a fracture becomes unstable it will slip, and the associated dilation slip will increase the fracture transmissivity. The transmissivity increase is introduced in a simplified manner: if a fracture segment reaches the slip condition, its transmissivity is multiplied by a fixed permeability enhancement factor. In reality, slip on a fracture perturbs local the stress state, potentially affecting the stability of other fractures. This process is not implemented in the code because the direction and amount of slip is ambiguous. However, stress change induced by thermal changes is implemented, not yet in a fully coupled way, but instead computing the thermal stresses and superposing thermal stresses onto the ambient stresses.\\
In addition to the processes shown in Figure \ref{fig1:concept}, THERMAID  properly accounts for gravity effects, internal pressure and heat sources, and the pressure- and temperature-dependence of fluid density and viscosity.   The corresponding equations of state are given by \citep{sun2008} for density and \citep{alshemmeri2012} for the viscosity of water.

To summarize, THERMAID is a thermo-hydraulic code for fractured media that accounts for mechanical stability of the fractures, slip-induced transmissivity increase and thermally induced stresses. In the following, we introduce the governing equations for fluid flow and heat transport, couplings between fracture and matrix, and the implemented fracture stability analysis. 

\subsection{Governing equations}
Flow in naturally fractured reservoirs is often described by the equations for nearly incompressible single-phase flow. We assume that the equations for nearly incompressible single-phase flow are valid in both matrix and the fractures. This simplification might not yield an adequate description of the flow in some fractured reservoirs where very large fracture apertures result in non-Darcian flow. The methodology presented here is however easily modifiable to extended Darcy flow models.  

The pressure equation, derived from continuity and total mass balance equations for single-phase fluid flow, is:  
\begin{equation}
\phi \left(\beta_f + \beta_r\right)\frac{\partial p}{\partial t} = \nabla \cdot \left[ \frac{k}{\mu}(\nabla p-\rho_fg)\right] + Q \label{eq1:pressure}
\end{equation}
where $\phi$ [-] is the porosity, $\rho_f$ $[\frac{kg}{m^3}]$ is the fluid density, $p$ [Pa] is the fluid pressure and $Q$ $[\frac{m^3}{s}]$ a source term. The compressibilities $\beta$ [Pa$^{-1}$] are denoted with the subscripts $f$ for fluid and $r$ for rock, respectively, $k$ $[m^2]$ is the permeability and $\mu$ [Pa$\cdot$s] the fluid viscosity. We consider only isotropic permeability $k$. Permeability is often linked to fracture aperture through the Cubic law, which has been shown to be useful in predicting fluid transport through fractured reservoirs and fractured porous media in general. However, it does not account for the roughness of the fracture or flow adjacent to the fracture walls due to the rock permeability. From the fluid pressure $p$, the fluid velocity is calculated using Darcy's law, i.e.
\begin{equation}
\mathbf{v} = - \frac{k}{\mu} \left[ \nabla p -\rho_f g\right]\label{eq1:darcy_v}
\end{equation}
The total mass balance equation derived above is separated into parts for the matrix and the fracture domains, i.e.
\begin{equation}
\phi^m\left( \beta_f + \beta_r\right)\frac{\partial p^m}{\partial t} = \nabla \cdot \left[ \frac{k^m}{\mu^m}(\nabla p^m-\rho_fg)\right] + \Psi^{mf} +Q^m  \label{eq1:pressure_m}
\end{equation}
and
\begin{equation}
\phi^{f} \left(\beta_f + \beta_r\right)\frac{\partial p^f}{\partial t} = \nabla \cdot \left[ \frac{k^f}{\mu^f}(\nabla p^f-\rho_fg)\right] + \Psi^{fm} +Q^f
\end{equation}
where $\Psi^{mf}$ and $\Psi^{fm}$ are the flux transfer functions between the matrix and the fractures. Superscripts $m$ and $f$ denote matrix and fracture quantities respectively.

The heat transport equation is derived similarly to the pressure equation based on a balance of energy. We assume local thermal equilibrium so that $T=T_r =T_f$ where $T_r$ and $T_f$ are the temperatures of solid rock and fluid respectively. Taking average over an elemental volume we find
\begin{equation}
\overline{c_p\rho}\frac{\partial T}{\partial t} + {c_p}_f\rho_f\mathbf{v}\nabla T - \overline{\lambda}\nabla^2 T = \overline{q} \label{eq1:heat}
\end{equation}
where over-lined properties denote volume averaged mean values for the porous medium.
\begin{eqnarray}
\overline{c_p\rho} &= \phi ({c_{p}}_f\rho_f) + (1-\phi)({c_{p}}_r\rho_r) \label{eq1:heat_cp} \\
\overline{\lambda} &= \phi \lambda_{f} + (1-\phi)\lambda_{r} \label{eq1:heat_lambda}\\
\overline{q} &= \phi q_{f} + (1-\phi)q_{r} \label{eq1:heat_q}
\end{eqnarray}
In equations \ref{eq1:heat} to \ref{eq1:heat_q} the heat capacity $c_{p} \,[\frac{J}{kg K}]$, the thermal conductivity $\lambda \,[\frac{W}{m K}]$ and internal heat source $q \,[\frac{W}{m^3}]$ of solid rock and fluid have been introduced. The fluid velocity $\mathbf{v}$ used in the heat transport equation is the Darcy velocity given by equation \ref{eq1:darcy_v}.

The heat transport equation is separated into matrix and fracture parts according to the same procedure as for the fluid pressure equation
\begin{equation}
\overline{c_p\rho}^m\frac{\partial T^m}{\partial t} + ({c_p}_f\rho_f\mathbf{v})^m\nabla T^m - \overline{\lambda}^m\nabla^2 T^m = \overline{q}^m + \Chi^{mf} 
\end{equation}
and
\begin{equation}
\overline{c_p\rho}^f\frac{\partial T^f}{\partial t} + ({c_p}_f\rho_f\mathbf{v})^f\nabla T^f - \overline{\lambda}^f\nabla^2 T^f = \overline{q}^f + \Chi^{fm} 
\end{equation}
where $\Chi^{mf}$ and $\Chi^{fm}$ are the heat transfer functions between the damaged matrix and the fractures.  
\subsection{Fracture matrix coupling}
To obtain a conservative set of equations, we apply a transfer function governing the mass and heat exchange between the two domains. The transfer function is treated as a source/sink term in the pressure and heat transport equations for damaged matrix and fracture, respectively, similar to classical well models \citep{peaceman1978}.\\
The transfer function for the pressure equation is defined as 
\begin{equation}
\Psi^{fm} = CI \cdot \Xi \cdot (p^f-p^m)
\end{equation}
with $\Xi$ being the mean total mobility of the fluid, defined as the fraction of permeability and viscosity \citep{lee2001}. $CI$ is the connectivity index between matrix and fracture that is dependent on the numerical discretization (cf. next section). From the separated mass balance equations, it becomes immediately clear that the total flux between matrix and fracture has to be conserved:
\begin{equation}
\int \Psi^{mf} dV = - \int \Psi^{fm} dA
\end{equation}
The transfer function for the heat equation is similarly defined. However, as two heat transport mechanisms are present in the equation, the transfer function needs to account for both mechanisms. Thus, the transfer function is defined as:
\begin{equation}
\Chi^{fm} = {\Chi^{fm}}^{\nabla} + {\Chi^{fm}}^{\nabla^2}
\end{equation}
where the superscript $\nabla$ denotes the heat advection contribution and $\nabla^2$ denotes the heat conduction contribution. The heat conduction contribution ${\Chi^{fm}}^{\nabla^2}$ is derived using the same approach as in the pressure transfer function.
\begin{equation}
{\Chi^{fm}}^{\nabla^2} = CI \cdot \Lambda \cdot (T^f-T^m)
\end{equation}
Here, $\Lambda$ is the heat conductivity at the fracture-matrix interface which can be calculated as
\begin{equation}
\Lambda = \frac{2\cdot \lambda^f\cdot \lambda^m}{\lambda^f + \lambda^m} \label{eq1:Xi}
\end{equation}
using the definition of the averaged heat conductivity $\lambda$ given in equation \ref{eq1:heat_lambda}. The advection contribution $\chi_{fm}^{\nabla}$, on the other hand, explicitly shows the coupling to the pressure equation based on the Darcy velocity:
\begin{equation}
{\Chi^{fm}}^{\nabla} = \Upsilon \cdot \mathbf{v}^{fm} \label{eq1:Chi_advection}
\end{equation}
In equation \ref{eq1:Chi_advection} we introduce the fluid velocity $\mathbf{v}^{fm}$ and specific heat capacity $\Upsilon$ at the matrix-fracture interface. $\Upsilon$ is calculated analogous to equation \ref{eq1:Xi} and based on the averaged specific heat capacity given in equation \ref{eq1:heat_cp}. The fluid velocity at the matrix-fracture interface is defined as 
\begin{equation}
\mathbf{v}^{fm}= - CI \cdot \Xi \cdot (\nabla p)^{fm} \label{eq1:v_fm}
\end{equation}
where $(\nabla p)^{fm}$ is the pressure gradient at the interface of matrix and fracture.
As discussed for the pressure transfer function also the heat transfer flux has to be conserved:
\begin{equation}
\int \Chi^{mf} dV = - \int \Chi^{fm} dA
\end{equation}
\subsection{Fracture stability}
Within THERMAID, a simplified analytical approach to fracture slip enables us to estimate fracture stability based on slip tendency analysis. Following Amonton's law for purely frictional fault reactivation 
\begin{equation}
\tau = \mu_s \cdot \sigma_{\text{n}eff} \label{eq1:amonton}
\end{equation}
with $\tau$ as shear stress, $\sigma_{\text{n}eff}$ as effective normal stress ($\sigma_{\text{n}} -p$ and $p$ as fluid pressure), and $\mu_s$ as sliding friction coefficient \citep{byerlee1978}, slip tendency is the ratio of shear stress to effective normal stress on a surface \citep{morris1996}, i.e. 
\begin{equation}
T_s = \frac{\tau}{\sigma_{\text{n}eff}}
\end{equation}
Fracture failure or slip is likely to occur if the shear stress to effective normal stress ratio equals or exceeds the frictional sliding resistance $\mu_s$. Thus we define the stability of a fracture as follows
\begin{equation}
T_s =
\begin{cases}
\frac{\tau}{\sigma_{\text{n}eff}} < \mu_s       & \quad \text{(stable)}\\
\frac{\tau}{\sigma_{\text{n}eff}} \geq \mu_s    & \quad \text{(unstable)}\\
\end{cases}
\end{equation}
Shear and effective normal stress acting on a given fracture depend on the orientation of the fracture plane within the effective principal stress field. If the effective principal stress field $(\vec{l},\vec{m},\vec{n})$ and the reference coordinate system $(\vec{x},\vec{y},\vec{z})$ of the simulation match, shear and normal stress can be calculated by simple expressions based on the dip angle of the fracture \citep[e.g.][]{miller2004}. Otherwise, stress transformations are needed in order to calculate the correct normal and shear stress in the fracture coordinate system $(\vec{u},\vec{v},\vec{w})$. The involved rotation matrices can be calculated if the orientations of the principal stresses are known \citep[e.g.][]{allmendinger2011}. Finally we compute the normal and shear stress on the fracture in the fracture coordinate system based on the 3D principal stresses. Note that in this calculation the influence of the intermediate principal stress component is taken into account despite the 2D model geometry.

An important addition to the effects of pore pressure and far field stresses for fracture stability is thermal stress. A body will change its shape and/or volume when exposed to a temperature change $\Delta T$. If the body's deformation is restricted, as it would be the case for a small volume inside a rock mass, the strain results in thermal stress. 
\begin{equation}
\sigma_{T_{ij}} = \frac{E}{1-2\nu} \cdot \alpha \Delta T \delta_{ij} \label{eq1:sigma_t}
\end{equation}
where $\alpha$ is the coefficient of linear thermal expansion in $\frac{1}{K}$, $E$ is the Young's modulus ($Pa$) and $\nu$ the Poisson ratio (-). $\delta_{ij}$ is the Kronecker delta, which is 1 for identical indices $i$ and $j$, and 0 otherwise. The thermal stress is positive (relative compression) if the temperature difference is positive ($\Delta T > 0$), and if the temperature difference negative, the thermal stress is negative (relative tension).
In the following we assume that the thermal stress is independent of the fluid pressure and the in-situ stress state of the rock. Thus, the resulting stress can be obtained by superposition of the effective stress ($\sigma_{eff} = \sigma_{tot} - p$) and the thermal stress. We formulate the superposed effective stress as
\begin{equation}
\sigma_{eff} = \sigma_{tot} - p + \sigma_{T}
\end{equation}
which can be used in equation \ref{eq1:amonton} in order to account for thermal stress during the fracture stability analysis.\\ 
Clearly, other stress contributions as slip induced stresses and stresses induced by chemical reaction have to be considered in a general case. However, especially the estimation of slip induced stress changes is ambiguous as the amount of slip and slip direction for potentially failing fractures is not known a-priori unless the underlying equations for fracture slip are solved explicitly. Thus, for reasons of simplicity we restrict ourselves to only effective stress and thermally induced stress changes.

As fractures are reactivated they generally show an increase in aperture as the fracture surfaces are not smooth but have many asperities. Due to a strong aperture dependence of permeability \citep[e.g.][]{nemvcok2002}, where small changes in aperture result in very large changes in permeability, it can be assumed that unstable (or sliding) fractures undergo a stepwise change in fracture permeability \citep{millernur2000}. Here we adopt the most simple model
\begin{equation}
k^{f} = \begin{cases}
k^{f} & \quad \text{if } T_s < \mu_s \\
\gamma \cdot k^{f} & \quad \text{if } T_s \geq \mu_s \\
\end{cases}
\end{equation}
where $\gamma$ is an permeability enhancement factor.  
This model successfully described the distribution of the induced seismicity in the Basel EGS site, and fluid-driven aftershock sequences \citep{miller2004,miller2015}. 
\section{Implementation}\label{sec1:implementation}
We implemented the two-dimensional embedded discrete fracture method in MATLAB. Our implementation is based on the concepts used in \textit{MAFLOT}, an open source MATLAB flow solver \citep{kuenze2012}.  As briefly discussed in the introduction, the matrix and fracture domains are discretized by regular Cartesian grids in 2D for the matrix and 1D for the fractures respectively (cf. Figure \ref{fig1:edfm_domain}).
\subsection{Numerical discretization in space}
Using a finite volume approach, we discretize the domain $\Omega$ as the integration over finite control volumes $\Omega_{ij}$ with $\Omega = \sum_{ij=1}^N \Omega_{ij}$. Using the Gauss theorem, the divergence integral over the volume can be rewritten as the surface integral normal to the boundary of the volume. Applied to a matrix grid cell on the right hand side (RHS) of the pressure equation \ref{eq1:pressure_m} this yields
\begin{equation}
\int_{\Omega_{ij}} \nabla\cdot \left(\frac{k}{\mu}\cdot\nabla p\right)^m + \Psi^{mf} + Q^m dV \Rightarrow  \int_{\partial\Omega_{ij}}\left(\left(\frac{k}{\mu}\cdot\nabla p\right)^m +\Psi^{mf}\right)\cdot \mathbf{n} ds + \int_{\Omega_{ij}} Q^m dV
\end{equation}
Note that gravity is neglected here and in the remains of this section to better facilitate  comprehension of the implementation. The pressure gradient over the cell boundary $\partial\Omega_{ij}$ is approximated by a two-point flux approximation that is second-order accurate in space. As the domain is generally heterogeneous in terms of rock properties, a harmonic averaging technique is used to calculate the appropriate values at the cell boundaries. 

The discretization of the  RHS of the temperature equation is analogous to the pressure equations and omitted here for brevity. It is  worth noting, however, that the advection term must be treated with special care. In this EDFM implementation, we use an upwind method in the fractures in combination with a \textit{minmod-}flux limited QUICK scheme in the matrix \citep{courant1952,leonard1979,roe1986}.

\subsection{Connectivity index}
The connectivity index $CI$ between matrix and fracture is discretization-dependent, and defined based on the linear pressure distribution assumed within a grid cell intersected by a fracture \citep{hajibeygi2011}. It is defined as the length fraction $A_{ij,k}$ of fracture segment $k$ inside matrix cell $ij$ divided by the average distance $\langle d \rangle_{ij,k}$ between matrix cell $ij$ and fracture segment $k$.
\begin{equation}
CI_{ij,k} = \frac{A_{ij,k}}{\langle d \rangle_{ij,k}}
\end{equation}
The average distance $\langle d \rangle_{ij,k}$ can be calculated as 
\begin{equation}
\langle d \rangle_{ij,k} = \frac{\int x_k(x')dx'}{V_{ij}} \label{eq1:d_mean}
\end{equation}
where $x_k$ is the distance from the fracture within the matrix cell and $V_{ij}$ the volume of the matrix cell. This allows a proper accounting for the reduced influence of a fracture segment on a matrix cell if the fracture segment does not cross the matrix cell through its center. In many cases equation \ref{eq1:d_mean} has to be evaluated by numerical integration. For rectangular grids however, there exists an analytical solution \citep{hajibeygi2011,pluimers2015}. For enhanced efficiency, the analytical expressions are used in our implementation.

\subsection{Fracture intersections}
Fracture intersections significantly impact flow dynamics in the reservoir. The additional fracture-fracture transmissivity can be calculated as
\begin{equation}
T_{i,j}= \frac{\alpha_i \cdot \alpha_j}{\alpha_i + \alpha_j} \quad \textrm{with } \alpha_i = \frac{b_i \Xi_i}{0.5\cdot \Delta x^f}
\end{equation}
where $b_f^i$ denotes the fracture aperture, $\Xi_i$ the total mobility and $\Delta x^f$ the numerical discretization spacing in the fracture \citep{karimi2003}.
\subsection{Time-discretization}
The time derivatives in equations \ref{eq1:pressure} and \ref{eq1:heat} are treated using the backward Euler method, which is an implicit time-discretization with local truncation error $\mathcal{O}(h^{2})$. The method is unconditionally stable theoretically allowing arbitrarily large time steps. In practice, when encountering non-linear behavior, such as the temperature- and pressure-dependent evolution of fluid density, issues with non-convergence might appear and place an indirect restriction on the time-step. Nonetheless, much larger time steps are allowed in the implemented method when compared to explicit schemes.
\subsection{Solution strategy}
We adapt a serial iterative scheme in order to accurately account for the coupling between the pressure and transport equations. In strongly coupled problems, multiple iterations must be used to capture any arising nonlinearities. In most cases, the flow and transport exhibit rather loose coupling in which only a few iterations are needed to converge to the solution. If on the other hand, fracture stability ceases and permeability enhancement in unstable fracture parts is used, the number of iterations might increase significantly and even limit the timestep.
\section{Results}\label{sec1:results}
We present the results of three benchmark experiments and an application experiment that provide insight into the capabilities of THERMAID and validate the implemented method. Fracture permeability and aperture are treated as independent from each other in the following. This allows simulating 'filled' fractures with relatively high aperture and comparably small permeability and allows fracture permeability estimates independent of Cubic law.

First we validate the implemented model with a simple flow problem independently. We then evaluate the coupled results of fluid flow and heat transport on a simple geometry and on a more realistic complex fracture network. The final numerical experiment is the application of the implemented approach to a field scale problem were we take advantage of the implemented fracture stability analysis in order to characterize the stimulated reservoir during injection of a geothermal reservoir.
\subsection{Validation of the pressure equation}
In order to validate the implementation of flow equations of the  model, we use an analytical solution for the steady-state flow in a porous medium in the presence of a fracture \citep{strack1982,kolditz2012}. Figure \ref{fig1:setup_H} shows the benchmark geometry, a square with a length of 10 m with a 2m-long inclined fracture in the center of a square domain. The aperture of the fractures is fixed at $b = 0.05m$. Uniform flow is maintained by imposing a specific discharge $q_0$ from the left boundary into the domain. To compare numerical results with the analytical solution, pressures calculated by the analytical solution are used at the lateral boundaries, i.e. $p_{in} = 49646$ Pa and $p_{out} = -49646$ Pa (cf. Figure \ref{fig1:setup_H}). On the top and bottom a no-flow Neumann boundary is applied. The remaining material properties of the numerical model are shown in Table \ref{tab:param}.

\begin{table}[!hb]
	\caption{Model parameters used for the inclined fracture solution.}
	\vspace{-0.5cm}
	\begin{center}
		\begin{tabular}{|l|l|l|l|}
			\hline
			& Parameter & Value & Unit\\ \hline
			$\alpha$ & Fracture angle & 45 & $^{\circ}$\\ %
			$b_{max}$ & Maximum fracture aperture & 0.05 & m\\ %
			$L $ & Fracture length & 2 & m \\ %
			$k^m$ & Matrix permeability & $1\cdot 10^{-12}$ & m$^2$ \\ %
			$k^f$ & Fracture permeability & $1\cdot 10^{-10}$ & m$^2$ \\ %
			$\mu$ & Fluid viscosity & $1\cdot 10^{-3}$ & Pa s \\ %
			$q_0$ & Specific discharge & $1\cdot 10^{-4}$ & m s$^{-1}$ \\ \hline
		\end{tabular}
		\label{tab:param}
	\end{center}
\vfill
	\centering
	\includegraphics[width=0.55\linewidth]{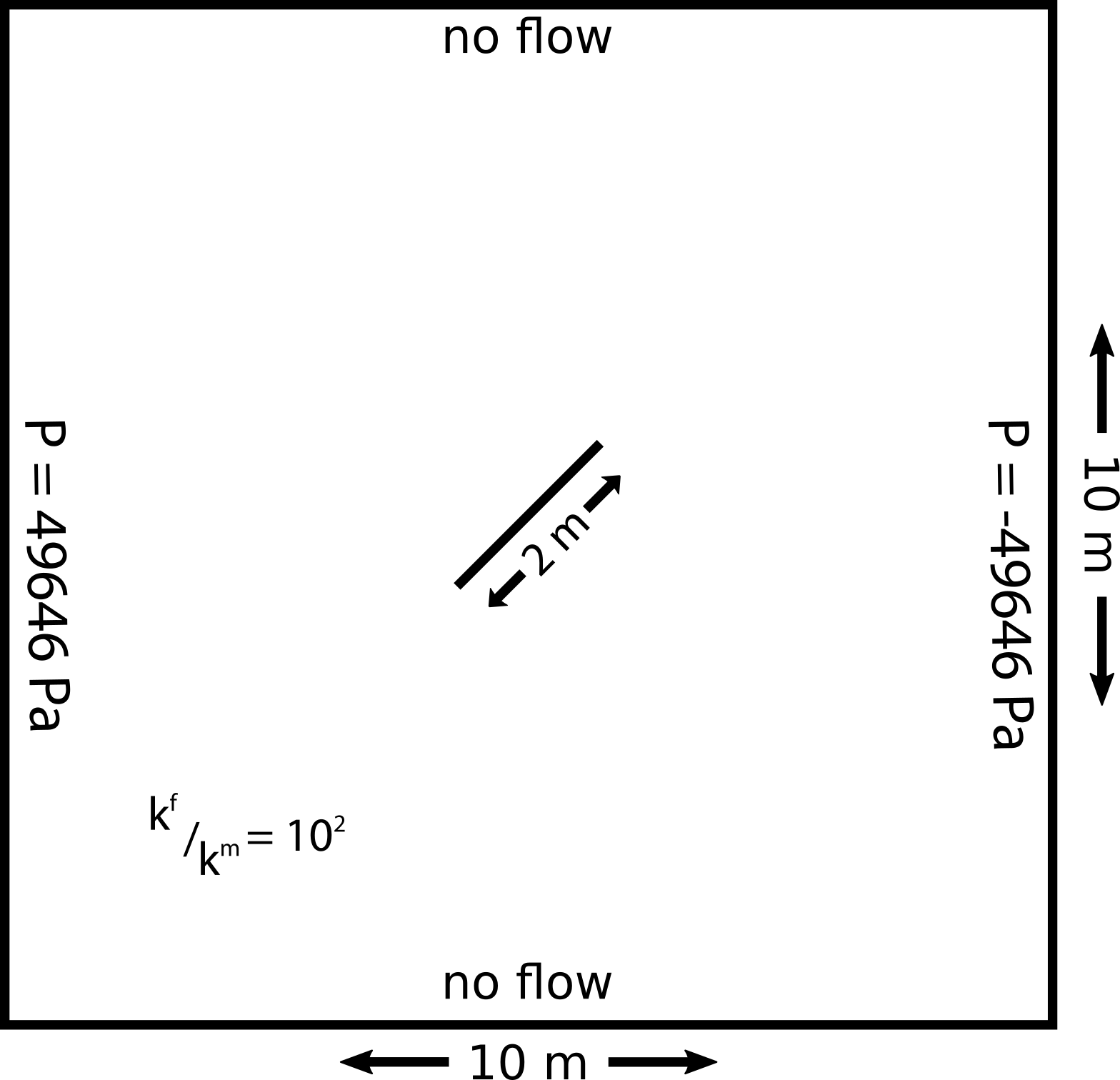}
	\caption{Numerical setup to evaluate the performance of the flow equations solution. Incompressible fluids are assumed in this benchmark experiment. On the left boundary a constant pressure of 49646Pa is assumed. The right boundary is set to -49646Pa to enforce the specific discharge $q_0$. On the other boundaries a no-flow boundary condition is applied.}
	\label{fig1:setup_H}
\end{table}

\begin{figure}[htbp]
  \centering
  \subfloat[]{ \includegraphics[width=0.44\textwidth]{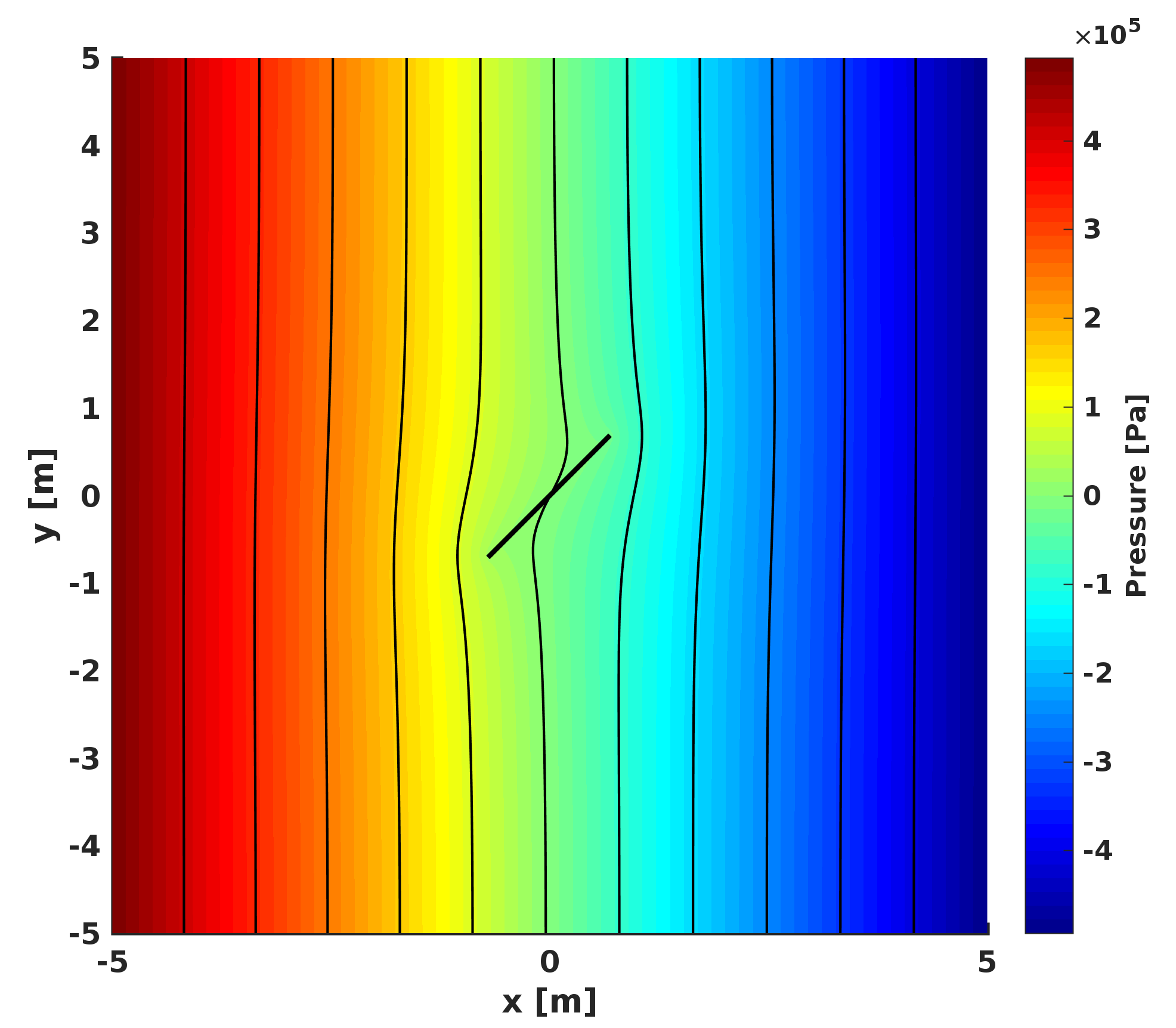}}
  \hfill
  \subfloat[]{\includegraphics[width=0.46\textwidth]{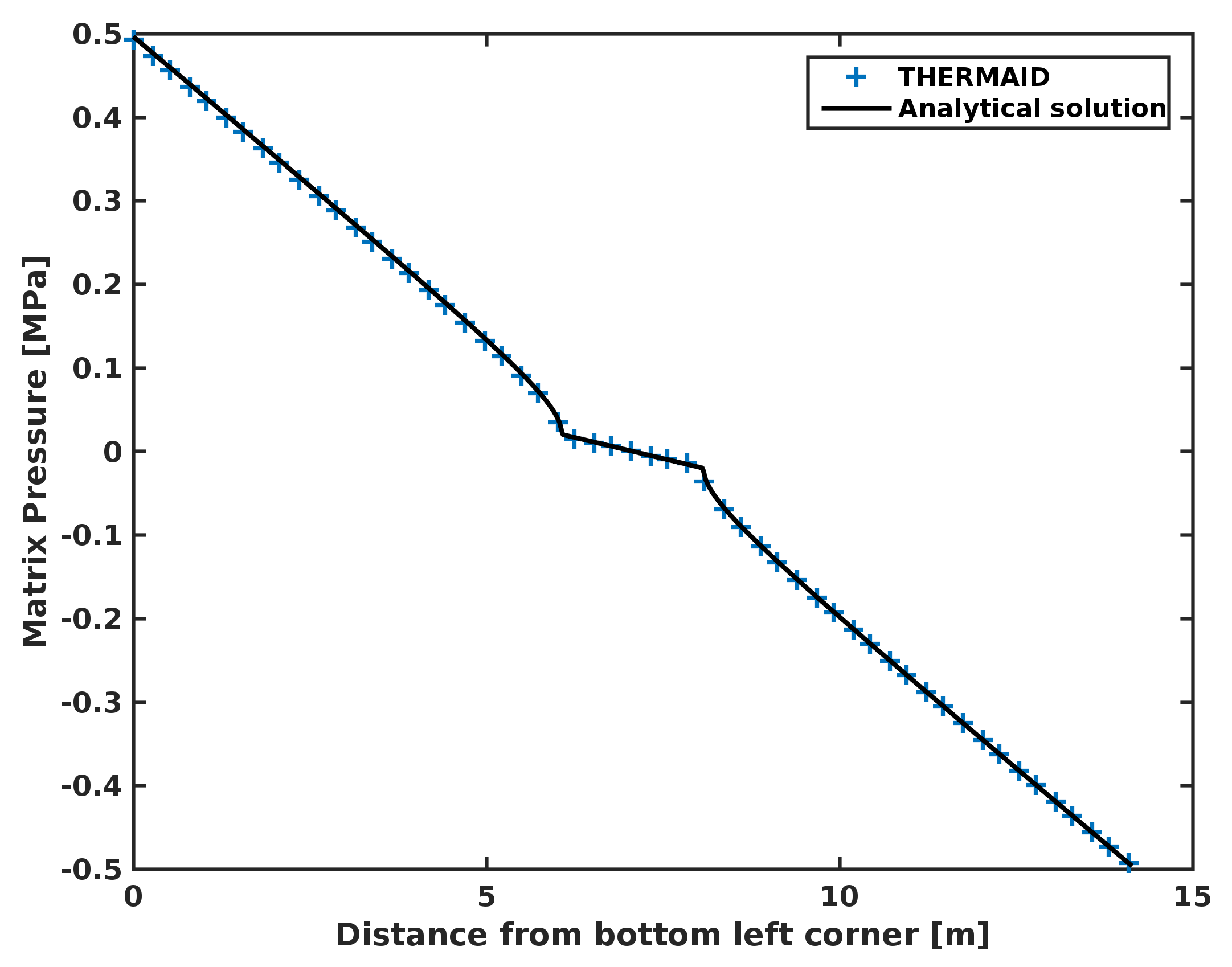}}
  \caption{a) Pressure field computed for the flow field including a single inclined fracture. b) Comparison between simulated (continuous red curve) and analytical derived (empty black circles) pressure distribution along diagonal from bottom left to top right of the model}
   \label{fig1:bench_H}
\end{figure}

The pressure distribution obtained by THERMAID is shown in Figure \ref{fig1:bench_H}a. The lateral uniform flow is disturbed in the vicinity of the inclined fracture where the flow is faster than in the surrounding porous media. Figure \ref{fig1:bench_H}b shows the pressure profile along a diagonal line from the bottom-left to the top-right. The results show very good agreement between the numerical solution obtained by THERMAID and the analytical solution. We quantify the difference between our model with the reference by the 'normalized root mean squared error (NRSME)' as well as the 'normalized mean absolute error (NMAE)'.
\begin{align}
NRSME &= \frac{\sqrt{\frac{\sum_{i=1}^n \left(x_i - x_i^{ref}\right)^2}{n}}}{max(x_i^{ref}) - min(x_i^{ref})} \\
\nonumber \\
NMAE &= \frac{\frac{\sum_{i=1}^n |x_i - x_i^{ref}|}{n}}{max(x_i^{ref}) - min(x_i^{ref})}
\end{align}
We decided to use two measures of performance due a recent debate on both measures \citep[e.g.][]{willmott2005,chai2014}. Especially \citep{chai2014} suggest that a combination of measures is required to assess model performance. We observe errors of well below $1\%$ (NRSME: 
$0.21\%$ and NMAE: $0.19\%$) that validate the implementation of the fluid flow equations.

\subsection{Validation of the heat transport equation}
\begin{figure}[!ht]
	\centering
	\vspace{0.25cm}
	\includegraphics[width=0.55\linewidth]{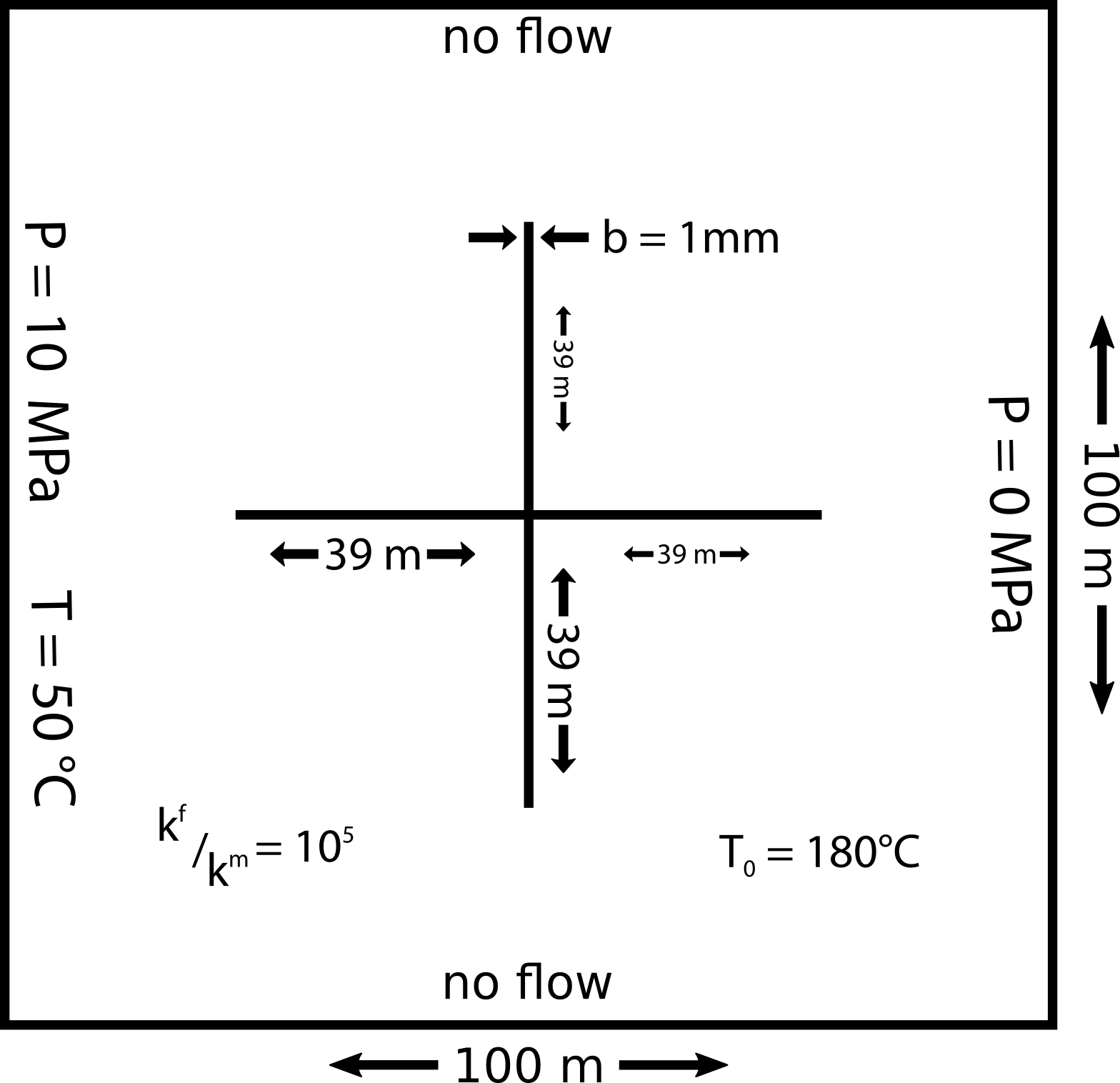}
	\caption{Numerical setup to evaluate the performance of the coupled flow and heat transport solution. A simple fracture geometry and incompressible fluids are used. On the left boundary a constant pressure of 10MPa at 50$^{\circ}$C is applied. The right boundary is set to 0Pa. On the other boundaries a no-flow boundary condition is applied. The interior has an initial temperature of 180$^{\circ}$C. All parameters for this model setup are shown in Table \ref{tab1:param_1}.}
	\label{fig1:setup_TH2}
\end{figure}

We validate the coupled flow and heat transport equations using a benchmark geometry that consists of two perpendicular 5m-long fractures intersecting  in the middle of a square domain (cf. Figure \ref{fig1:setup_TH2}). The aperture of the fractures is fixed at $b = 1mm$. The domain is 100m by 100m square domain with Dirichlet boundary conditions on the left and right sides. On the left a constant pressure of 10MPa is applied, whereas the right side is fixed to 0MPa. On the top and bottom a no-flow Neumann boundary is applied. The domain is initially at $T_0=180^{\circ}C$, which is a typical temperature for economic heat extraction in a geothermal reservoir. The inflow temperature at the left side of the domain is set to $T_{in}=50^{\circ}C$ (cf. Figure \ref{fig1:setup_TH2}). The material parameters for this benchmark were chosen realistically and are shown in Table \ref{tab1:param_1}. The benchmark's results are evaluated after 40 years of simulation. The matrix domain is discretized by 301x301 cells while the fractures are modeled by 304 fracture segments (152 each). The reference solution is computed by \textit{COMSOL} on a conforming discrete fracture network with a high resolution grid.

\begin{table}[!htbp]
	\caption{Properties used in the coupled flow and heat transport models. Superscripts: f - fracture, m - matrix. Subscripts: f - fluid, r - rock.}
	\begin{center}
		\begin{tabular}{|l|l|l|}
			\hline
			Permeability & $k^{f} = 1\cdot 10^{-11}m^2$ & $k^m = 10^{-16} m^2$\\ \hline
			Porosity & $\phi^{f} = 0.3$ & $\phi^m = 0.3$\\ \hline
			Density & $\rho_{f} = 1\cdot 10^{3}\frac{kg}{m^3}$ & $\rho_{r} = 2.5\cdot 10^{3}\frac{kg}{m^3}$\\ \hline
			Viscosity & $\mu_{f} = 1\cdot 10^{-3}Pa\cdot s$ & \\ \hline
			Specific heat & $c_{p_f} = 4000 \frac{J}{kg\cdot K}$ & $c_{p_r} = 1000 \frac{J}{kg\cdot K}$\\ \hline
			Heat conductivity & $\lambda_f = 0.5\frac{W}{m\cdot K}$ & $\lambda_r = 2.0\frac{W}{m\cdot K}$\\ \hline
		\end{tabular}
		\label{tab1:param_1}
	\end{center}
\end{table}

\begin{figure}[!ht]
	\centering
	\includegraphics[width=0.92\linewidth]{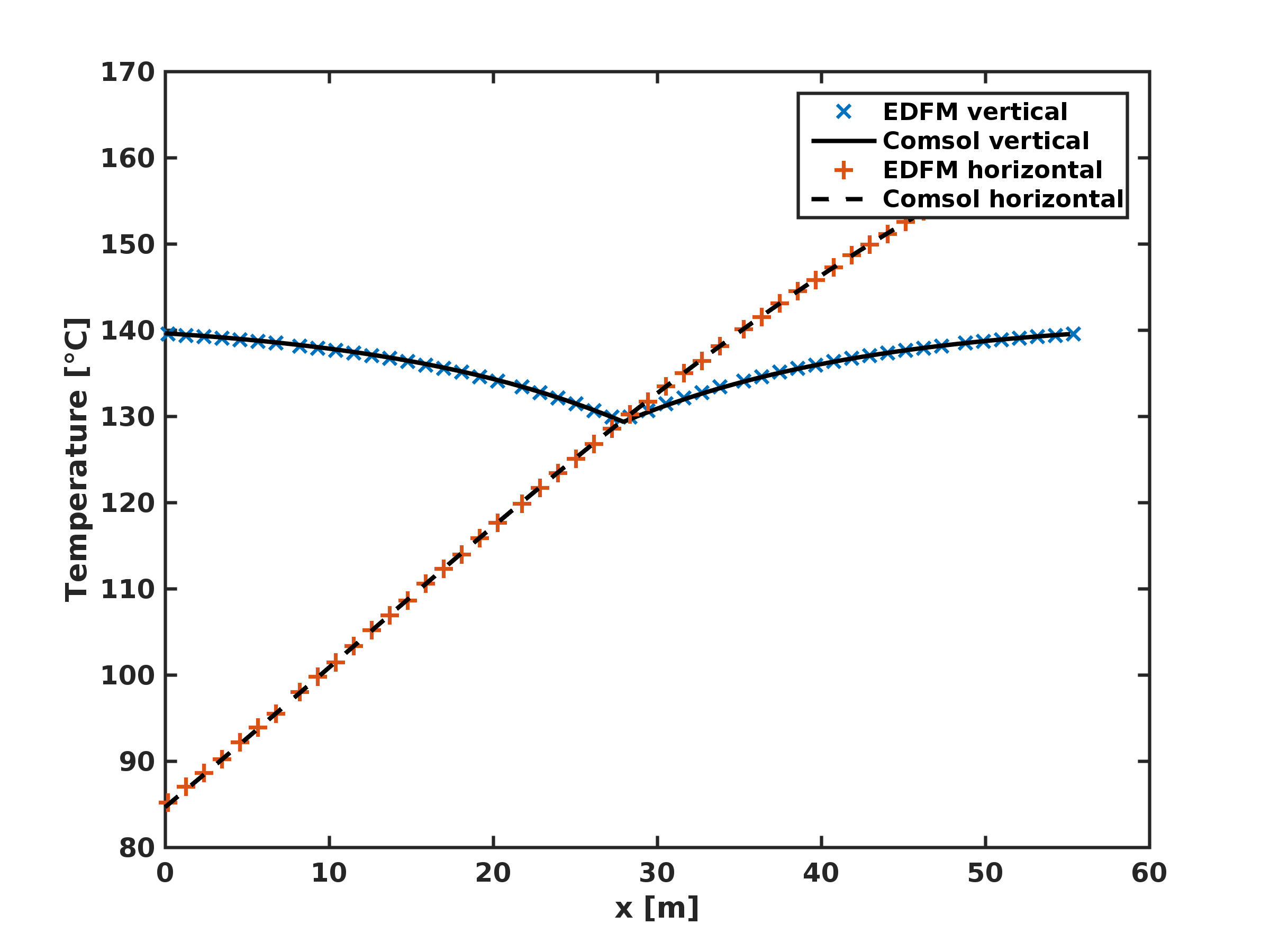}
	\caption{Fracture temperatures through vertical and horizontal fractures. For both fractures we see very good agreement between the implemented method and the reference solution.}
	\label{fig1:TH2_fracture}
\end{figure}

Figure \ref{fig1:TH2_fracture} shows the temperature in both fractures after 40 years of coupled flow and heat transport simulation. Additionally, Table \ref{tab1:TH2_error} shows the quantitative error analysis for the fracture temperatures. We observe a very good agreement between the temperature distribution in both fractures with the reference solution. The horizontal fracture presents changes in temperature over most its extent, which is in accordance with the principal flow direction. As the vertical fracture is not aligned with the flow, a rather homogeneous temperature decrease is observed to about $140^{\circ}C$ after 40 years. This is in good agreement with the matrix temperatures at the position of the fracture. Nevertheless, a significant change in temperature is observed close to the intersection of both fractures. Here the fracture-fracture interaction is clearly visible as both fractures show nearly identical temperatures at the intersection (cf. Figure \ref{fig1:TH2_fracture}). The quantitative error analysis shows differences between our solution and the reference of $\sim 0.8\%$ for the vertical fracture and $\sim 0.1\%$ for the horizontal fracture although the two error measures differ slightly (cf. Table \ref{tab1:TH2_error}). 

\begin{table}[!hb]
	\caption{NRSME and NMAE errors for the first coupled fluid flow and heat transport equation benchmark.}
	\begin{center}
		\begin{tabular}{|r|l|l|}
			\hline
			& $\mathbf{T_{vertical}}$ & $\mathbf{T_{horizontal}}$ \\ \hline
			\textbf{NRMSE [\%]}& $0.89$ & $0.16$\\ \hline
			\textbf{NMAE [\%]}& $0.71$ & $0.11$ \\ \hline
		\end{tabular}
		\label{tab1:TH2_error}
	\end{center}
\end{table}

Ultimately, the benchmark shows that our model accurately solves the coupled flow and heat transport equations for this geometry. The simulated time-frame is consistent with the estimated lifetime of a typical enhanced geothermal reservoir and additionally shows that the implemented time-marching scheme is accurate for the problem at hand.
\subsection{Validation of the heat transport equation on a complex fracture network}
\begin{figure}[!htbp]
	\centering
	\includegraphics[width=0.5\linewidth]{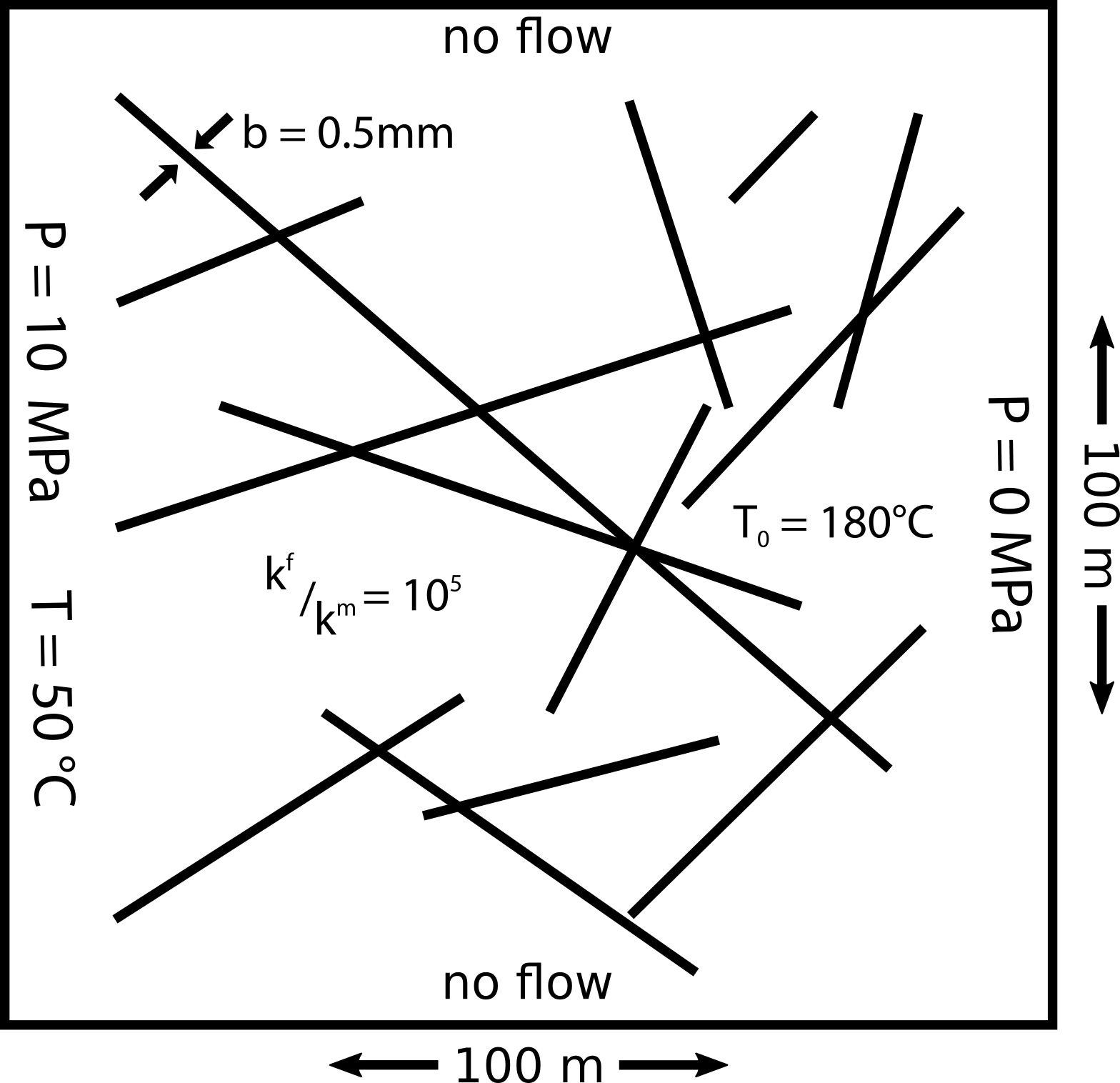}
	\caption{Numerical setup to evaluate the performance of the coupled flow and heat transport solution with a more realistic complex fracture geometry. Incompressible fluids are used. On the left boundary a constant pressure of 10MPa at 50$^{\circ}$C is applied. The right boundary is set to 0Pa. On the outer boundaries a no-flow boundary condition is applied. The interior has an initial temperature of 180$^{\circ}$C. All parameters for this model setup are shown in Table \ref{tab1:param_1}.}
	\label{fig1:setup_TH}
\end{figure}

We evaluate the coupled flow and heat transport on a more complex fracture geometry. The geometry (Figure \ref{fig1:setup_TH}) consists of a total of 13 fractures within a square domain. Boundary and initial conditions are equal to the previous experiment. The fracture aperture is set to $b = 0.5mm$. The remaining parameters governing the heat transport are consistent with the benchmark in the last section and shown in Table \ref{tab1:param_1}. We evaluate the results after 40 years of simulation. The reference solution computed by \textit{COMSOL} contains 419'594 DOF. In this experiment we evaluate also grid dependence of the implemented model by comparing the results for different resolution simulations with the reference.

\begin{figure}[htbp]
	\centering
	\subfloat[]{\includegraphics[width=0.49\textwidth]{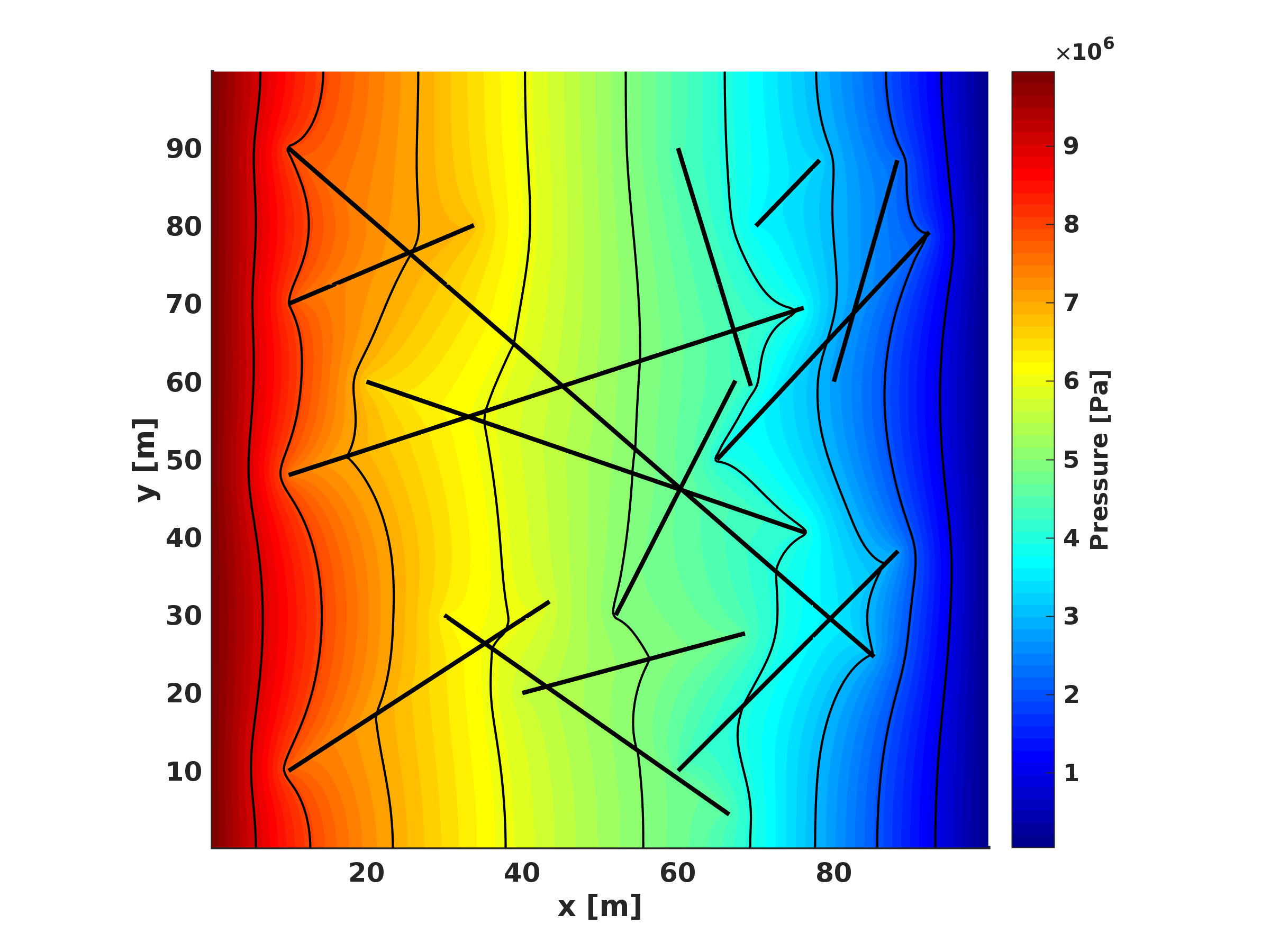}}
	\hfill
	\subfloat[]{ \includegraphics[width=0.49\textwidth]{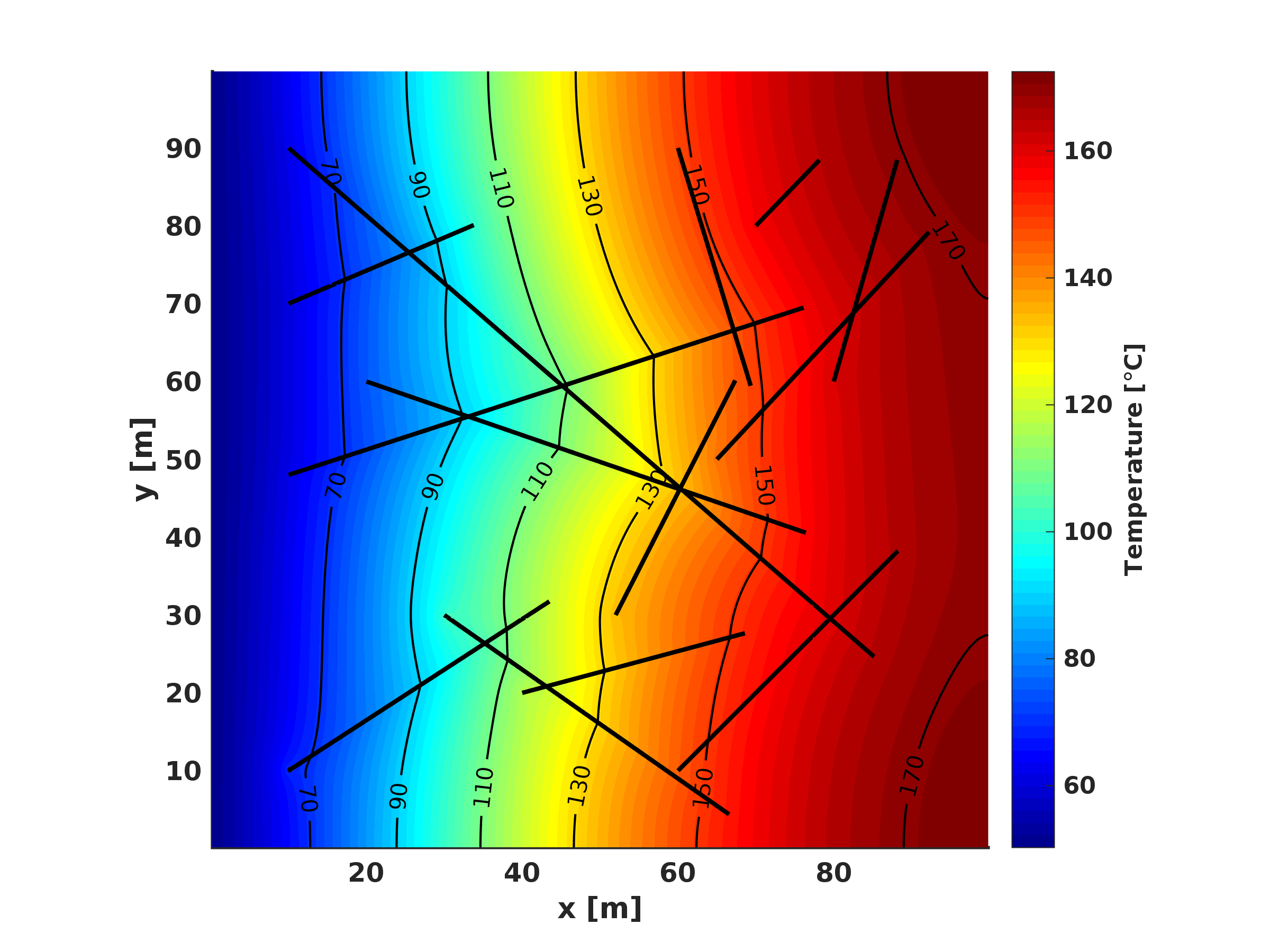}}
	\caption{a) Pressure distribution for the complex fracture geometry. The heterogeneous pressure distributions shows the significant influence of the fractures. b) Final temperature distribution in the matrix for the complex fracture geometry. The heterogeneous pressure distributions leads to inhomogeneous fluid velocities, which is consequently shown in the temperature evolution.}
	\label{fig1:TH_matrix}
\end{figure}

In the previous section we focused on the temperature distributions in the fractures. Here we take a closer look at the matrix temperature distributions. Figure \ref{fig1:TH_matrix}a shows the final pressure distribution for a matrix grid resolution of 301x301. The temperature distribution in the domain after 40 years of simulation is shown in \ref{fig1:TH_matrix}b. Both pressure and temperature fields show a heterogeneous distribution due to the influence of the fractures. 

\begin{figure}[htbp]
	\centering
	\subfloat[]{\includegraphics[width=0.49\textwidth]{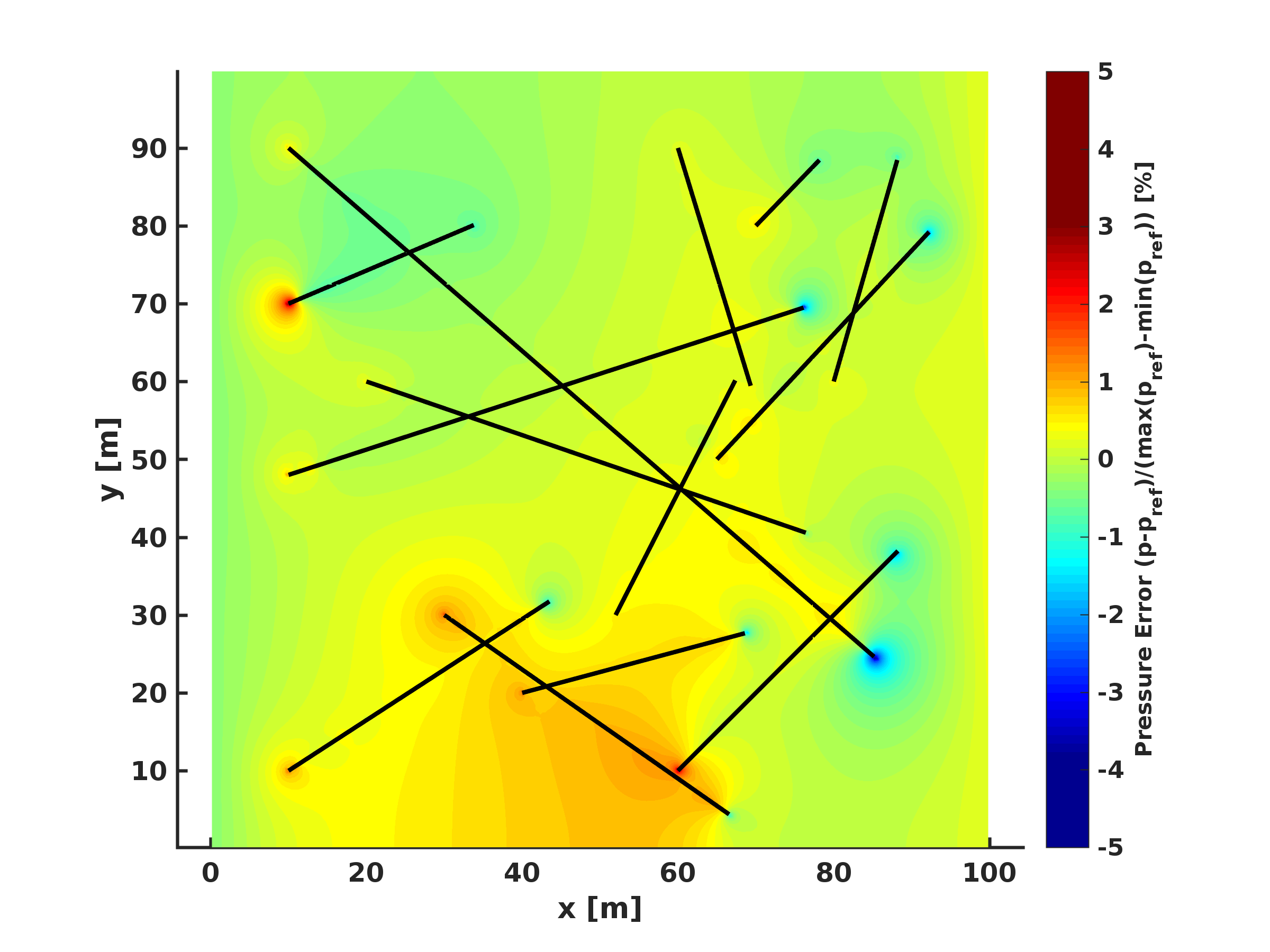}}
	\hfill
	\subfloat[]{ \includegraphics[width=0.49\textwidth]{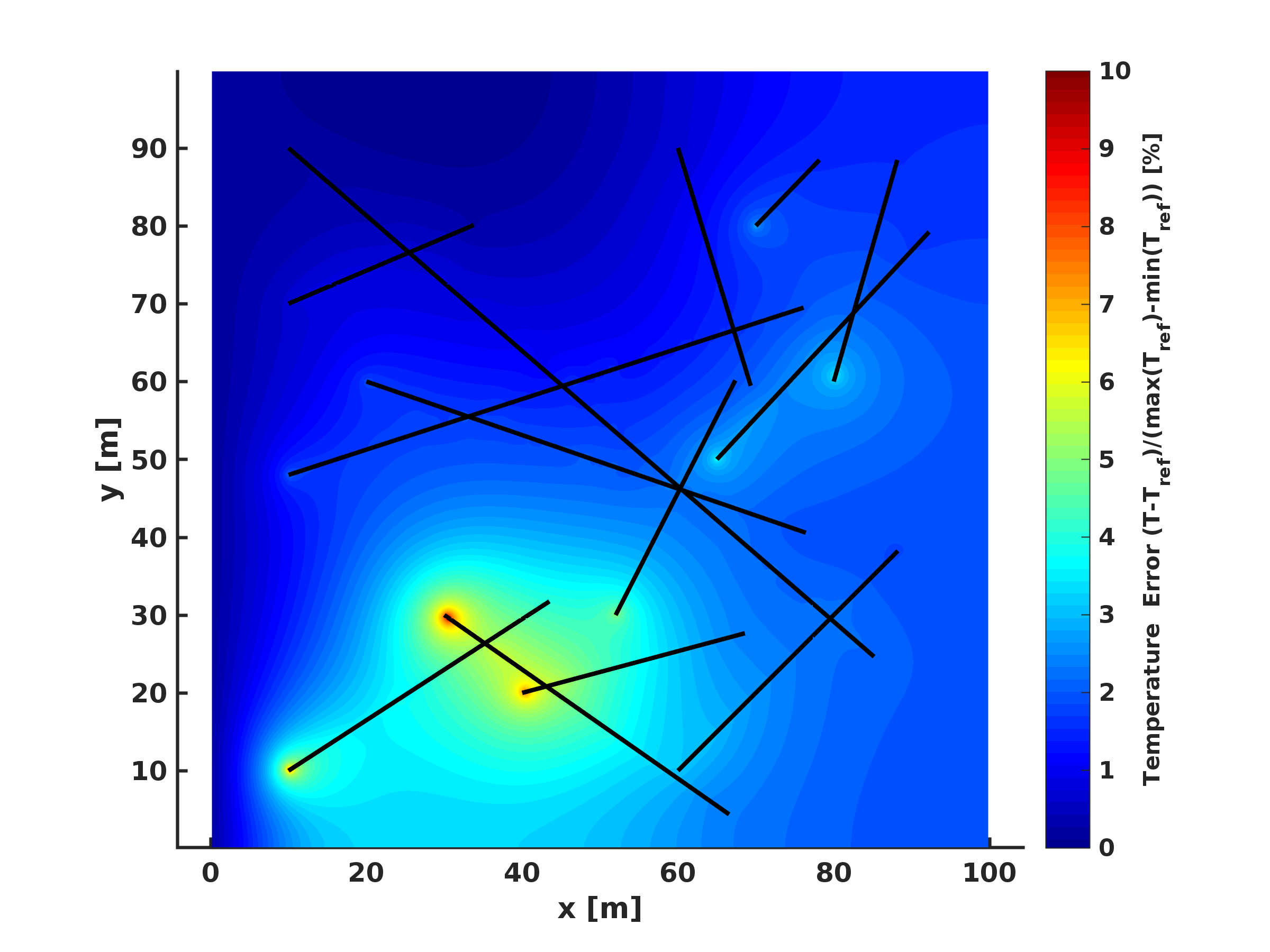}}
	\caption{Deviation from reference solution: \textbf{a)} Pressure. A region of elevated error is visible in the lower part of the domain. Significant deviations are also visible at some fracture tips. \textbf{b)} Temperature. Measured at the final stage of the simulation. The error in the pressure solution are reproduced in the temperature solution. The matrix temperatures are generally overestimated compared with the reference.}
	\label{fig1:TH_error}
\end{figure}

Figures \ref{fig1:TH_error}a and \ref{fig1:TH_error}b show the percental deviation of our solution from the reference for the matrix grid resolution of 301x301 of our model.  The pressure solution shows only small errors with a NRMSE of $0.35\%$. In the lower third of the domain between 20m and 60m in $x$-direction, a region of elevated error ($\sim 1\%$) is present (cf. Figure \ref{fig1:TH_error}a). Bigger deviations are visible close to some fracture tips where typically on the high-pressure (inflow) side of the fracture our model overestimates the matrix pressure compared to the reference. The low-pressure (outflow) sides of the fractures show predominantly underestimations of pressure. Interestingly, fractures that exhibit error concentration around one of the tips, do not necessarily show the opposite error on the other side of the fracture. Maximum pressure deviations from the reference are below $\pm 5\%$.  \\
The errors in the temperature distribution are generally larger than for the pressure. Figure \ref{fig1:TH_error}b shows the percentage error at the final stage of the simulation for a matrix grid discretization of 301x301. Compared with the error in the pressure solution, we find that our model seems to always overestimate the matrix temperature compared to the reference. The normalized RMS error for this resolution is $2.22\%$. We suspect that the elevated temperature deviations are caused by the relatively small error in the pressure solution. The small error in the pressure leads to comparably larger differences in flow velocities that are controlling heat advection. Thus, over a simulation of 40 years this error accumulates to the values observed here. 

We further want to investigate the influence of the resolution on the accuracy of the results, so we compare four different resolutions (101x101, 301x301, 501x501 and 1001x1001). We investigate the improvement of solution accuracy in the pressure and heat transport solutions by using the NRMSE and NMAE values compared to the high resolution solution obtained by \textit{COMSOL}. Table \ref{tab1:TH_error_resolution} shows both error measurements for all resolutions. We find a general improvement of the accuracy with an increase in resolution. For the temperature, this is a decrease in NRMSE from $4.6\%$ (101x101) to $1.2\%$ (1001x1001). The pressure error is consistently about one magnitude smaller, showing a decrease from $0.78\%$ (101x101) to $0.15\%$ (1001x1001). Overall we find a significant increase in accuracy with an increase in resolution. Nevertheless, the deviation is not changing significantly between 501x501 and 1001x1001 ($1.74\%$ vs $1.22\%$ in case of the temperature). This indicates a systematic difference between the reference solution and our method. There are multiple possible origins of this systematic error. Since we observe the systematic deviation also in the pressure, we think it is likely to be a difference in methodology concerning the fluid flow equation. These differences could include the treatment of fracture-fracture intersections, the definition of matrix-fracture interface permeability, and inherent numerical differences between finite element and finite volume methods. Nevertheless, we find very good agreement between the reference simulation and our implementation for large parts of the model. Even in regions of significantly elevated deviation, we find acceptable agreement with differences below $10\%$ between the two methods. The definitive source of the difference is currently not resolved but presents excellent future research opportunities.

\begin{table}
	\caption{NRSME and NMAE errors for the second coupled fluid flow and heat transport equation benchmark in dependence of resolution.}
	\begin{center}
		\begin{tabular}{|r|l|l|l|l|}
			\hline
			& $\mathbf{p^m}$ & & $\mathbf{T^m}$ & \\ \hline
			& \textbf{NRMSE [\%]}& \textbf{NMAE [\%]}& \textbf{NRMSE [\%]}& \textbf{NMAE [\%]} \\ \hline
			\textbf{101 x 101} & 0.78 & 0.64 & 4.59 & 4.0 \\ \hline
			\textbf{301 x 301} & 0.35 & 0.26 & 2.22 & 1.87  \\ \hline
			\textbf{501 x 501} & 0.23 & 0.17 & 1.74 & 1.51 \\ \hline
			\textbf{1001 x 1001} & 0.12 & 0.12 & 1.11 & 1.02  \\ \hline
		\end{tabular}
		\label{tab1:TH_error_resolution}
	\end{center}
\end{table}

\subsection{Utilization of the fracture stability analysis}
\begin{figure}[!htbp]
	\centering
	\includegraphics[width=0.65\linewidth]{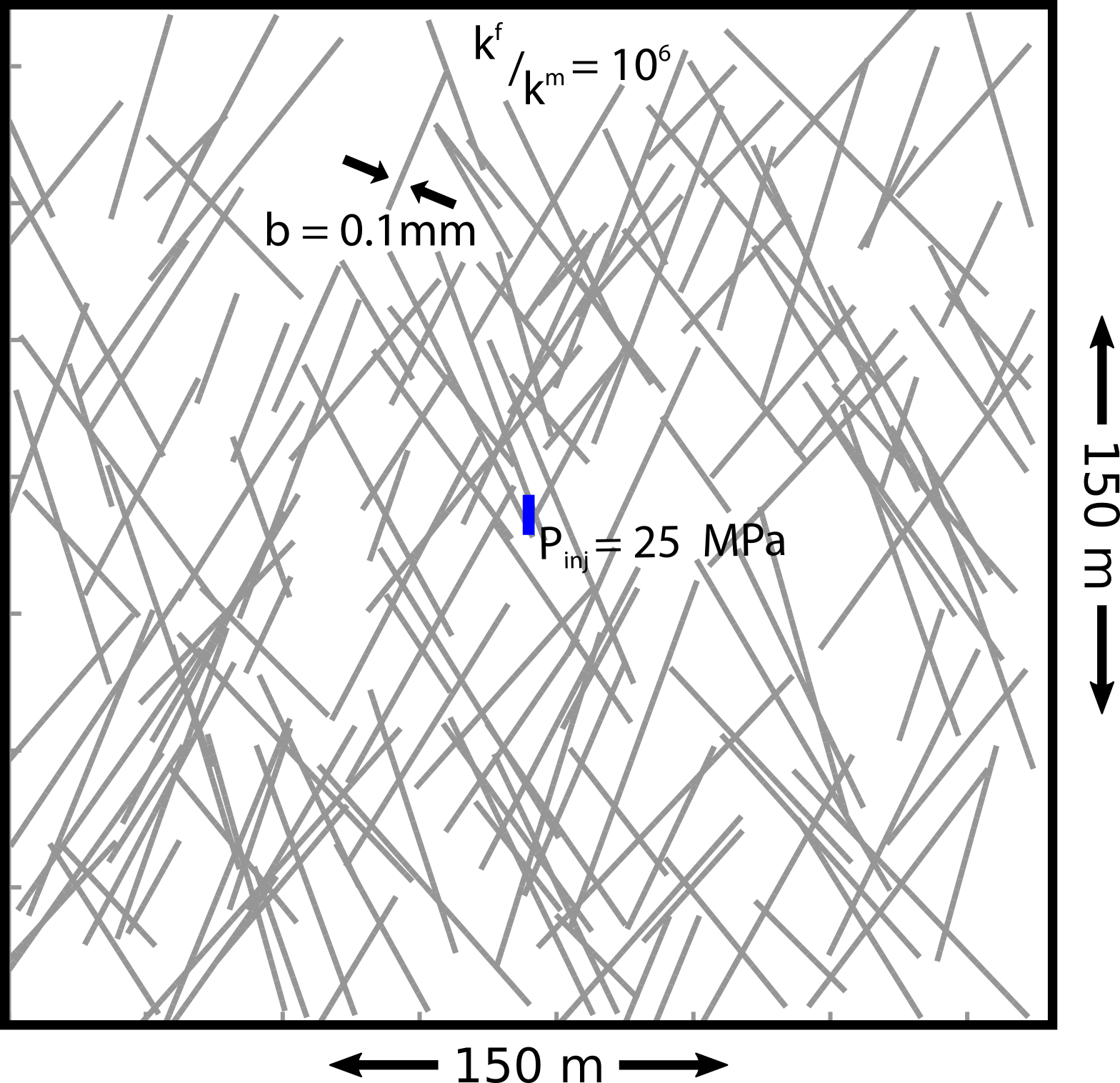}
	\caption{Numerical setup to evaluate the fracture stability on a realistic complex fracture geometry. A constant injection pressure of 25MPa is applied in the borehole (blue line). On the outer boundaries a no-flow boundary condition is applied. All parameters for this model setup are shown in Table \ref{tab1:param_2}.}
	\label{fig1:setup_FSA}
\end{figure}

\begin{figure}[!htbp]
	\centering
	\includegraphics[width=0.90\linewidth]{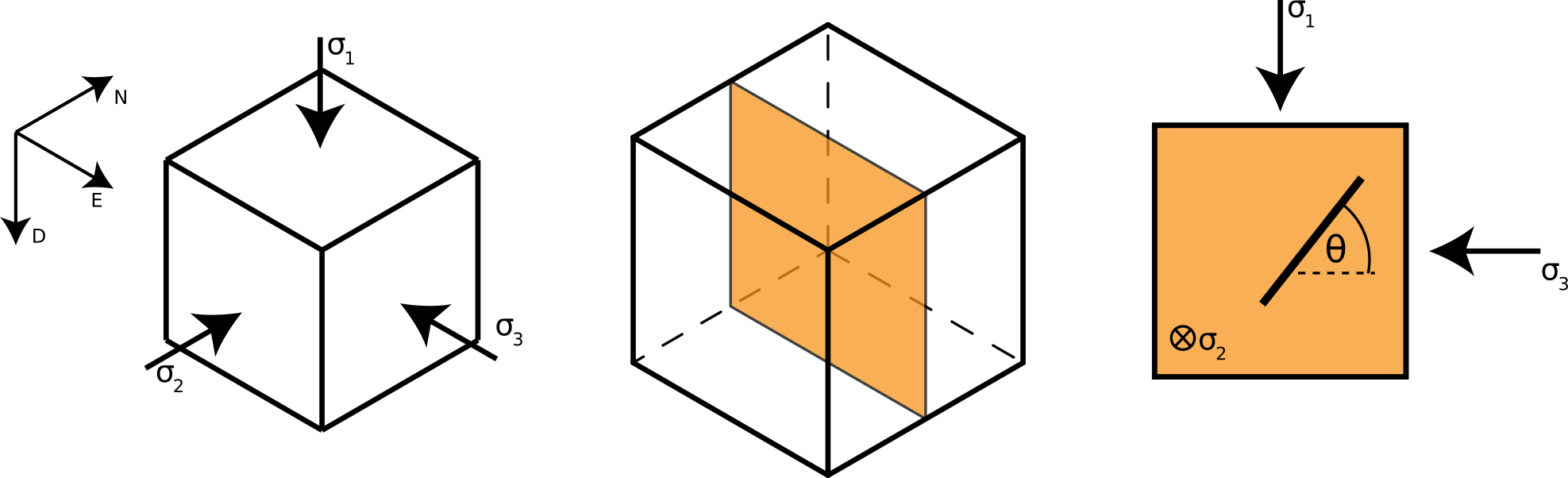}
	\caption{Principal stress orientations in the fracture stability analysis. The principal stress field is aligned with a NED coordinate system. The orientation of the reference plane (simulation plane) within the principal stress field is shown as well as the resulting stress orientations in 2D view.}
	\label{fig1:stress_systems}
\end{figure}

We present the fracture stability analysis to show the influence of permeability enhancement and thermal stress on fracture stability. We model  fluid injection into a complex fracture network with a range of fracture orientations. The geometry consists of a total of 196 fractures within a square domain (Figure \ref{fig1:setup_FSA}). The borehole is located in the middle of the domain with an open hole section of 6m. The fracture aperture in the reservoir is set to $b = 0.1mm$. The remaining parameters used in this section are shown in Table \ref{tab1:param_2}. The upper limit of the fractured reservoir domain is assumed to be at 5km depth. The injection pressure is held constant at 25MPa. The principal stresses are oriented as shown in Figure \ref{fig1:stress_systems}, which corresponds to a normal faulting regime. The magnitudes of the principal stresses are 125MPa, 107.5MPa and 81.25 MPa respectively, which corresponds to a relative stress ratio of $R=0.4$. The in-situ pore pressure is assumed to be hydrostatic ($\sim 50$MPa). The stress conditions roughly resemble the relative conditions at the Fenton Hill and Hijiori geothermal projects although both projects were situated above 4km depth \citep{xie2015,barton1988,oikawa2000}. We evaluate the results after 10 days of continuous fluid injection.

\begin{table}[!htb]
	\caption{Properties used in the fracture stability analysis model. Superscripts: f - fracture, m - matrix. Subscripts: f - fluid, r - rock.}
	\begin{center}
		\begin{tabular}{|l|l|l|}
			\hline
			Permeability & $k^{f} = 1\cdot 10^{-12}m^2$ & $k^m = 10^{-18} m^2$\\[5pt] \hline
			Porosity & $\phi^{f} = 0.9$ & $\phi^m = 0.1$\\[5pt] \hline
			Density & $\rho_{f} = 1\cdot 10^{3}\frac{kg}{m^3}$ & $\rho_{r} = 2.5\cdot 10^{3}\frac{kg}{m^3}$\\[5pt] \hline
			Viscosity & $\mu_{f} = 1\cdot 10^{-3}Pa\cdot s$ & \\[5pt] \hline
			Specific heat & $c_{p_f} = 4000 \frac{J}{kg\cdot K}$ & $c_{p_r} = 1000 \frac{J}{kg\cdot K}$\\[5pt] \hline
			Heat conductivity & $\lambda_f = 0.5\frac{W}{m\cdot K}$ & $\lambda_r = 2.0\frac{W}{m\cdot K}$\\[5pt] \hline
			Thermal expansion. coeff.& $\alpha = 7.9\cdot 10^{-6} K^{-1}$ & \\[5pt] \hline 
			Shear modulus& $G = 29.0 GPa$ & \\[5pt] \hline
			Poisson ratio & $\nu = 0.25$ & \\[5pt] \hline
		\end{tabular}
		\label{tab1:param_2}
	\end{center}
\end{table}

\begin{figure}[!htbp]
	\centering
	\includegraphics[width=0.9\linewidth]{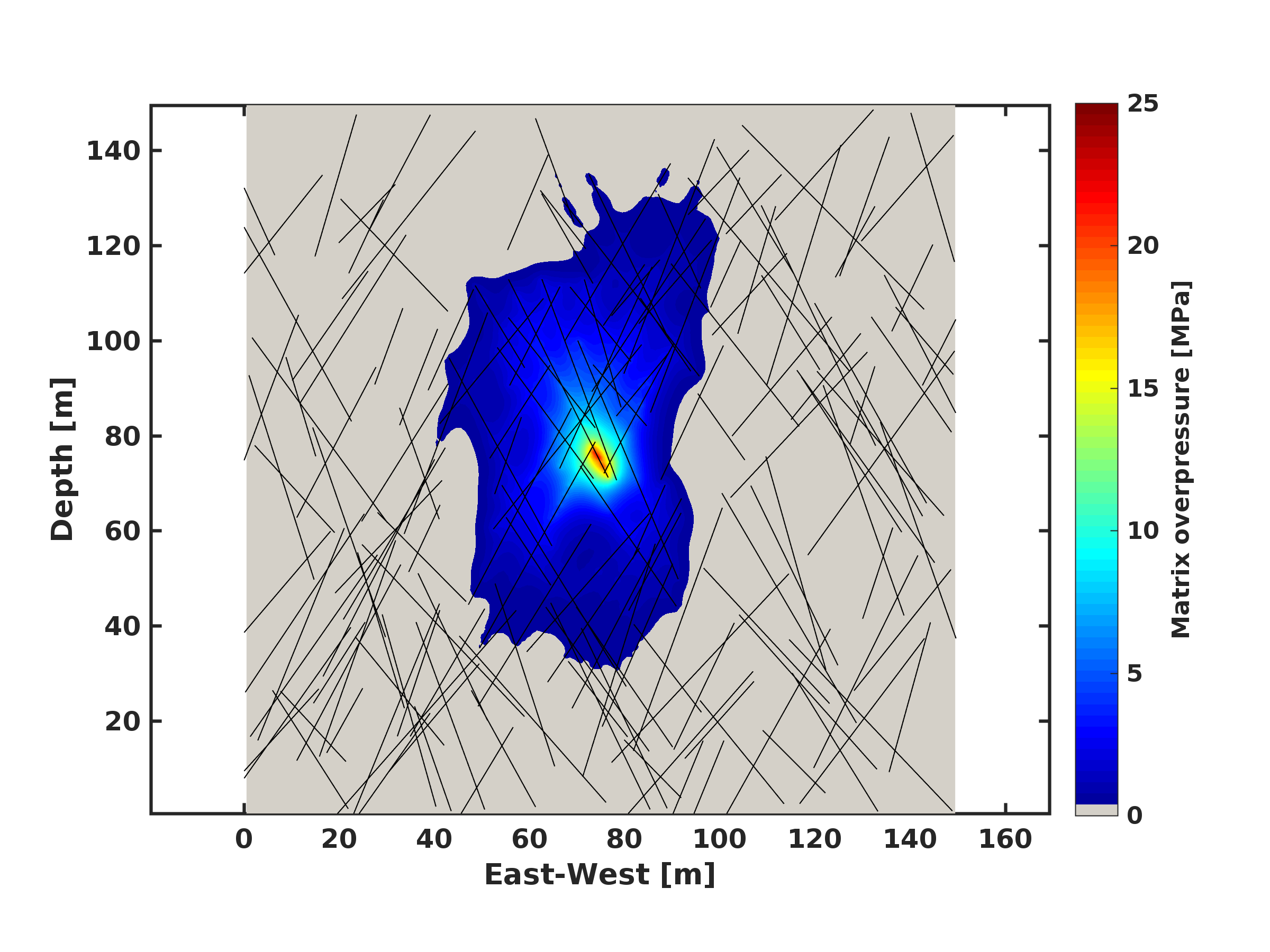}
	\caption{Matrix pressure in the reservoir after 10 days of injection. Due to the orientation of the fractures a preferential flow direction in the vertical direction is visible.}
	\label{fig1:FSA_pressure_1}
\end{figure}

\begin{figure}[!ht]
	\centering
	\includegraphics[width=0.9\linewidth]{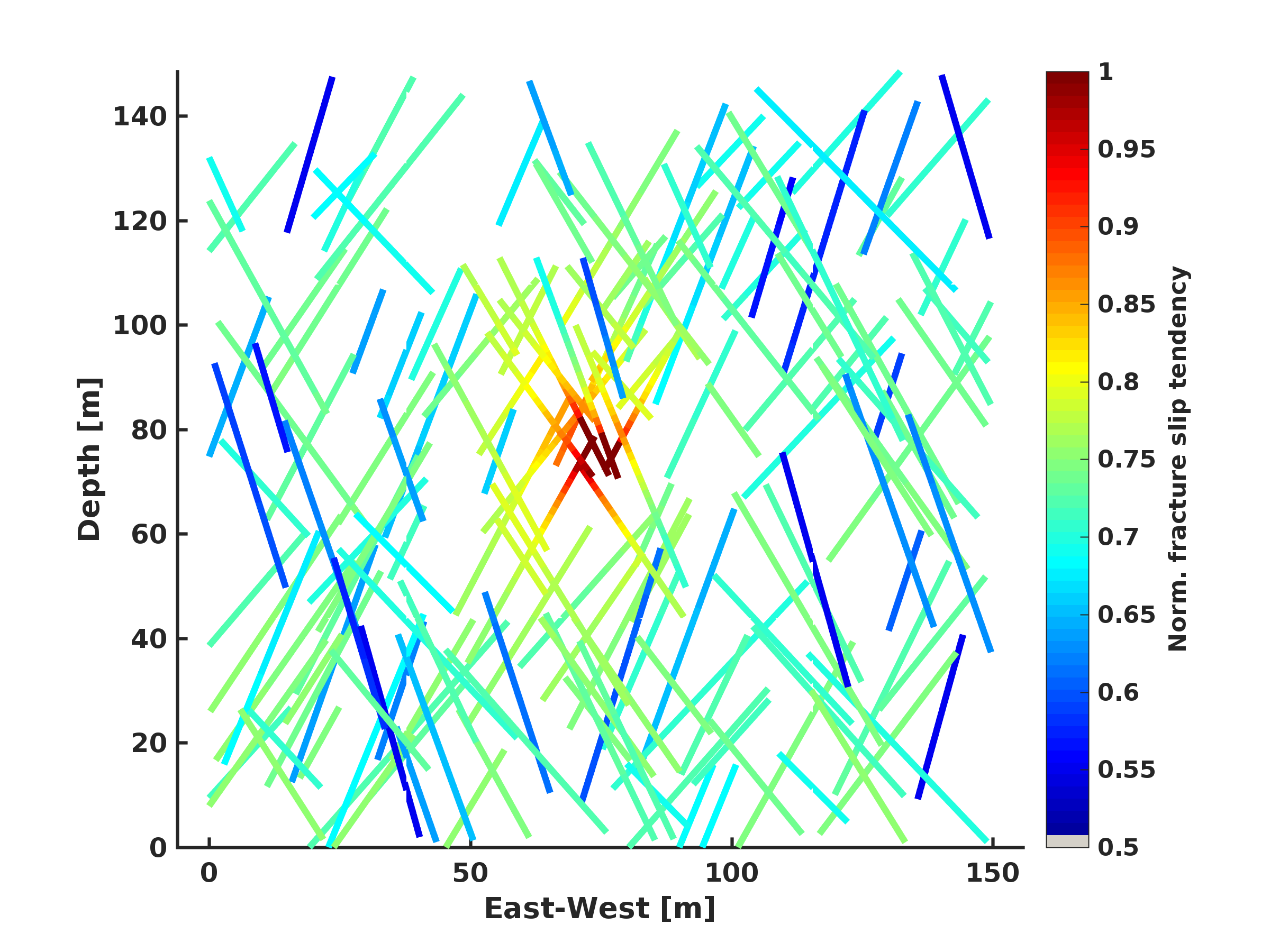}
	\caption{Fracture stability in terms of normalized slip tendency in the reservoir after 10 days of injection. Values are normalized by the friction coefficient $\mu=0.6$. High values denote higher slip tendency. The general fracture stability is good at levels well below the instability condition. Stability reduces closer to the injection. Very close to the injection point high fluid pressures lead to unstable fracture segments (red). }
	\label{fig1:FSA_1}
\end{figure}

Figure \ref{fig1:FSA_pressure_1} shows the pressure distribution after 10 days of injection. Due to the orientation of the pre-existing fractures, a preferential flow direction in the vertical direction is visible. Slight pressure changes due to the injection are measured at distances up to 55m in the vertical and 35m in the horizontal directions from the injection point. The zone of 10MPa pressure changes extends roughly 10m around the borehole. Very high pressures $>20$MPa are concentrated in the direct vicinity of the injection.\\ 
The in-situ fracture stability is influenced by the additional injected fluid pressure. Figure \ref{fig1:FSA_1} shows the final normalized fracture slip tendency. Note that a normalized slip tendency value of 1 represents a fracture that is eligible for failure and slip. We observe a range of values in the reservoir based on the fractures' orientations. The average fracture stability is high with values well below the failure condition. However, closer to the injection the increased slip tendency due to the injection is visible. Zones with fluid overpressure of $>5$MPa show significant increase in slip tendency (yellow colors in the plot). The region with at least 10MPa additional fluid pressure is very close to or eligible for slip on the fracture.

\begin{figure}[!htb]
	\centering
	\includegraphics[width=0.9\linewidth]{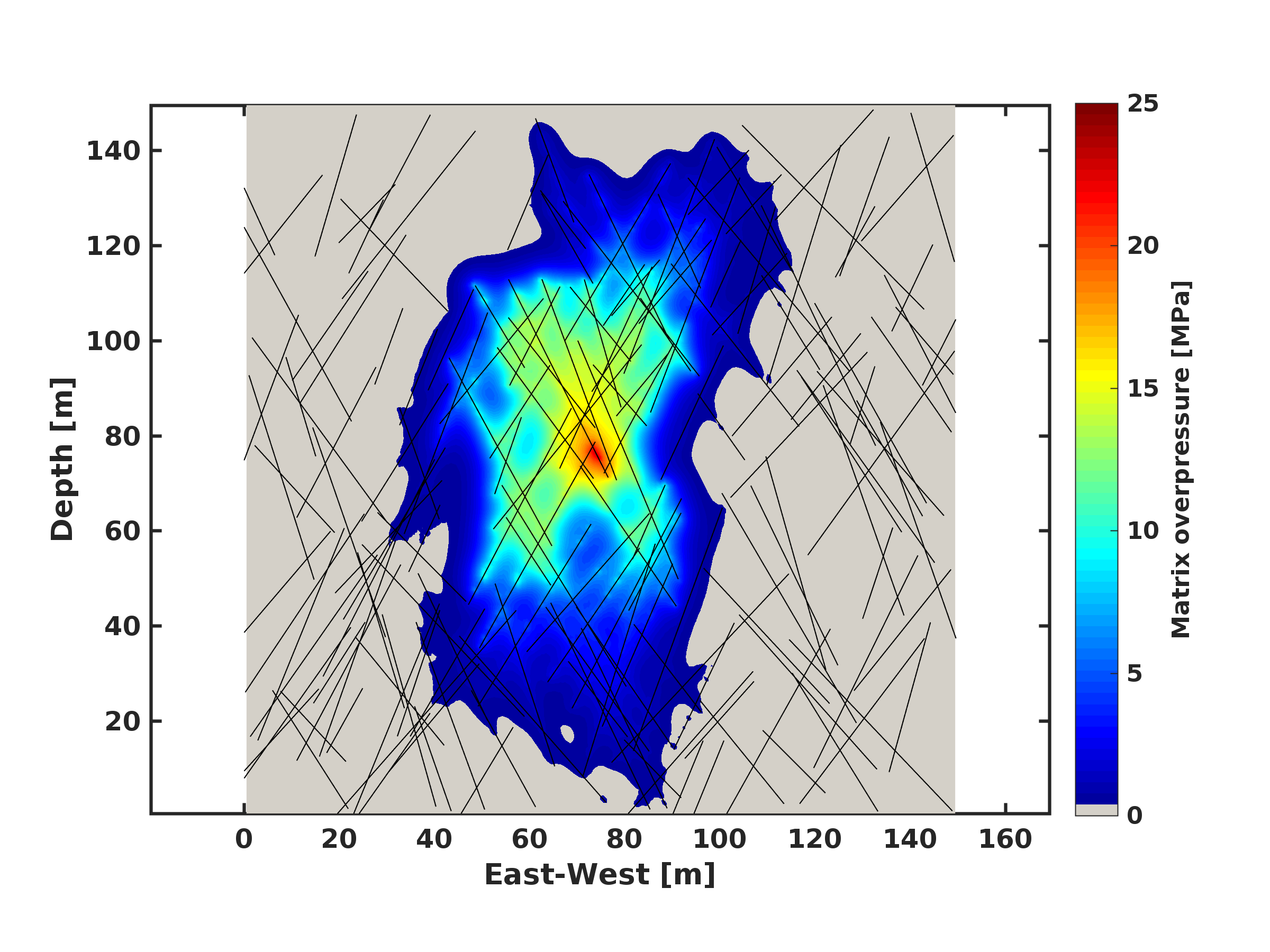}
	\caption{Matrix pressure in the reservoir after 10 days of injection if permeability enhancement is used. Here a enhancement factor of $\gamma=10$ is used. Due to the constant injection pressure and increased permeability in unstable fracture segments, the high-pressure zone is increased drastically.}
	\label{fig1:FSA_pressure_2}
\end{figure}

\paragraph{Permeability enhancement} 
In the previous section fracture segments eligible for slip did not have any feedback on the fluid pressure distribution. Here we investigate this feedback by introducing the stepwise permeability enhancement for failing fractures. The setup used is identical to the previous section except a 10-fold increase in permeability is assumed for failing fracture segments.\\ 
Figure \ref{fig1:FSA_pressure_2} shows the pressure distribution after 10 days of injection if permeability enhancement is considered. Although the general flow directions remain unchanged, the fluid pressure distribution shows significant differences in extent and magnitude.

\begin{figure}[!htb]
	\centering
	\subfloat[]{\includegraphics[width=0.49\linewidth]{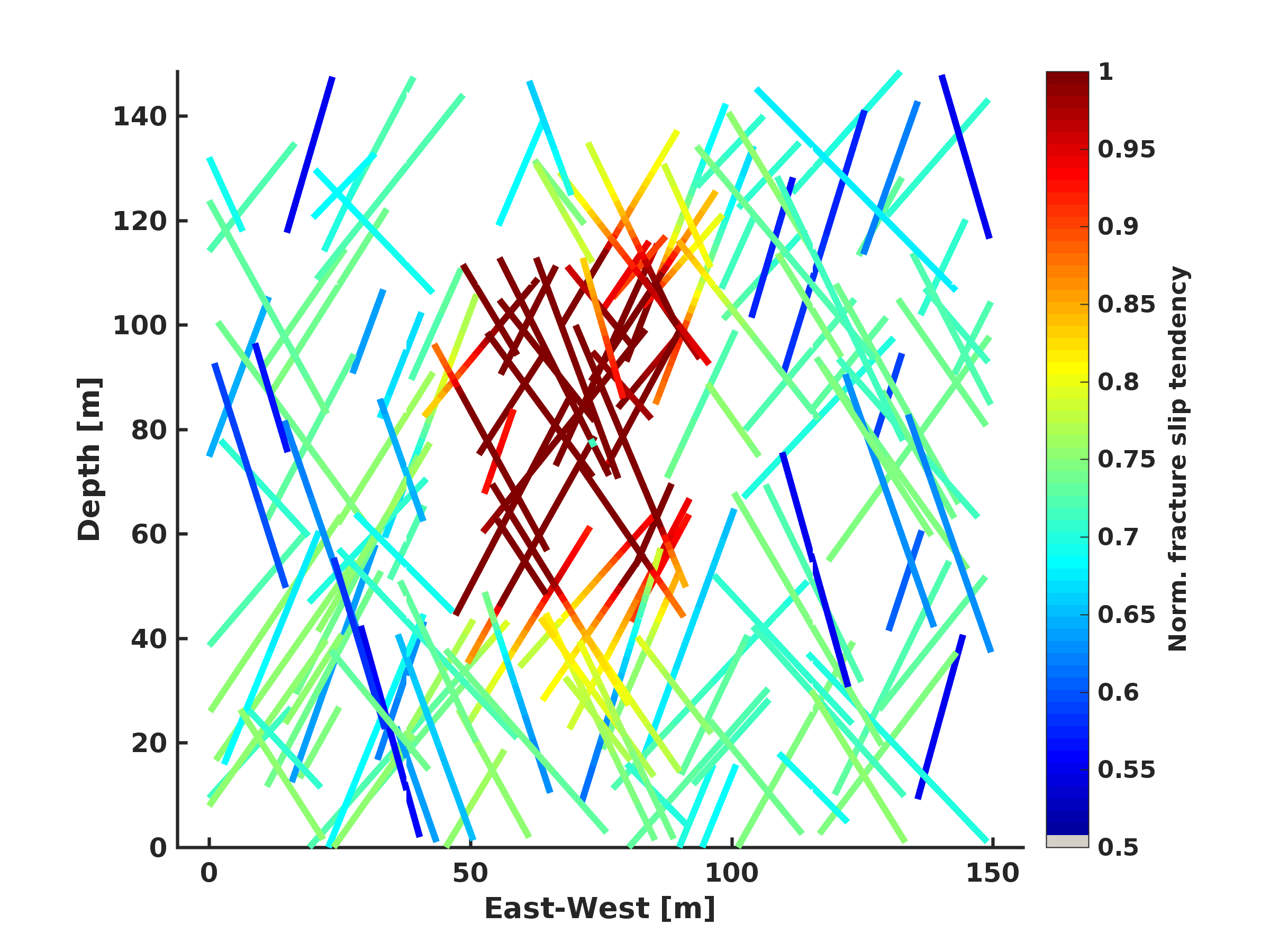}}
	\hfill
	\subfloat[]{\includegraphics[width=0.49\linewidth]{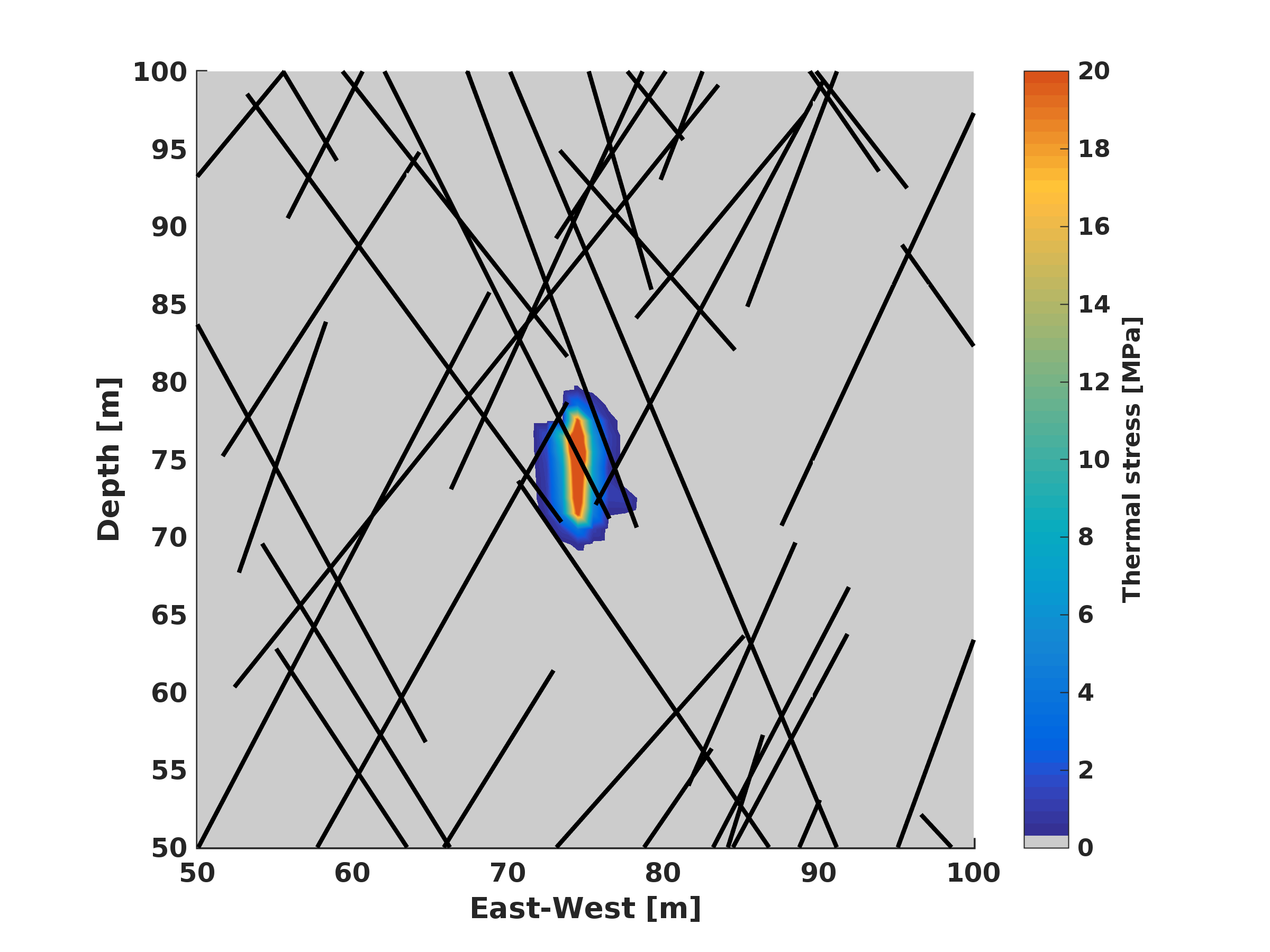}}
	\caption{\textbf{a)} Fracture stability in terms of normalized slip tendency in the reservoir after 10 days of injection for the case of permeability enhancement. Values are normalized by the friction coefficient $\mu=0.6$. High values denote higher slip tendency. Stability drastically reduces closer to the injection as the high fluid pressure zone is much bigger if permeability enhancement is used (red). \textbf{b)} Thermal stress after 10 days of injection into the fracture network. The thermal stress is concentrated close to the injection well. The color-scale in the figure starts at 0.25MPa with the darkest blue. Everything below is neglected in the graphical representation and shown in the background color. Note that the absolute value of the thermal stress is shown and all thermal stress here is tensional.}
	\label{fig1:FSA_2}
\end{figure}

Fracture stability is changed drastically if permeability enhancement is considered. The in-situ fracture stability remains unchanged in the outer regions of the domain at very stable levels. On the other hand, most of the fractures within the overpressured regions show elevated slip tendency with fractures closer to the injection being eligible for slip. Compared to the previous simulation, 20-times more fracture segments are unstable and capable of slip. Failing fractures, which have increased in permeability allow fluid to propagate more easily. As we assume a constant pressure injection, the amount of injected fluid is increased significantly. In this way a much larger stimulated area is observed compared to the case without permeability enhancement.\\ 
Thermal stress has only a small influence during the relatively short injection period of 10 days in this simulation. The resulting thermal stress distribution is shown in Figure \ref{fig1:FSA_2} and shows thermal stresses concentrated at the borehole. As the fractures within the vicinity of significant thermal stress are eligible for slip also by the injection fluid pressure, no additional unstable fracture segments are observed. However, in a recent study investigating the role of thermal stress in a geothermal reservoir in detail we found that thermal stresses can facilitate slip on non-optimally oriented fractures, and this is especially important in long-term injection scenarios where the thermal stress changes become more significant with time \citep{jansen2017}. \\ 

The experiments presented here show the importance of fracture stability analysis. We showed that a stepwise permeability increase in potentially failing fracture segments has a major impact on the stimulated reservoir volume and allows fracture slip in larger parts of the domain. This emphasizes the importance of coupling thermo-hydraulic models with the mechanical changes during fracture slip.

\section{Conclusion}\label{sec1:conclusion}
We developed, implemented and validated a fractured reservoir modeling framework in \textit{MATLAB} for investigating coupled thermo-hydraulic problems including fracture stability analysis. 

Our results show with high confidence that the accuracy of the implemented \textit{MATLAB} package are within the limits of commercial simulators for fractured reservoirs. Especially the results of the coupled flow and heat transport on a complex fracture network show the importance of discrete fractures in numerical analysis of fractured reservoirs. Both pressure and temperature distributions show heterogeneities due to fracture-matrix interactions. THERMAID presents easy access to the underlying implementation that enables rapid prototyping as well as detailed investigations of the embedded discrete fracture model and coupled processes in naturally fractured reservoirs. \\ 
As discussed earlier in the results, the deviations in the pressure solution could be caused by different treatment of fracture-fracture intersections or the definition of matrix-fracture interface permeability between the models. Currently there is no clear indication about which weighting to use at fracture-matrix interfaces, which presents an excellent future research opportunity for combined laboratory and numerical experiments. We assume that differences in the temperature solutions are caused by the deviations in the pressure solution that are magnified with time.

We showed that the embedded discrete fracture model is a viable alternative to the existing methods. As numerical discretization is simplified compared to conforming discrete fracture models, dynamic changes of the fracture network are possible without large numerical overhead. The extension of the embedded discrete fracture model to three dimensions has not been discussed so far in this article. Due to the relatively simple numerical discretization the extension to three dimensions is feasible. However, THERMAID is currently only developed in a 2D version. This is however not a limitation of the embedded fracture model but due limiting factors of the achievable computational performance in \textit{MATLAB}. Nevertheless, the approach taken in THERMAID could be efficiently re-implemented and extended to 3D in a high-performance computing environment. The embedded discrete fracture model is not necessarily restricted to regular grids and can be extended to general geometries. However, using regular grids can be advantageous for the application of massively parallel computation techniques to further increase computational efficiency and enable large scale, high resolution simulations.\\ 
Our results show the importance of including the mechanical behavior of fractures and the reservoir in thermo-hydraulic simulations. Although the deformation process during fracture slip was not explicitly taken into account, the assumed step-wise increase in fracture permeability during slip provides the necessary feedback for the pressure equation in order to capture the observed increase in injectivity during hydraulic stimulation. We propose that changes in permeability and aperture should be incorporated in all models that seek to fully understand the thermo-hydraulic evolution during fluid injection in fractured reservoirs. Although our model exhibits a very simplified view on the complex fracture mechanics, it still provides important insight into reservoir stimulation that helps in identifying some challenges and opportunities for future studies. More advanced models currently under development will consider both pre-existing fractures as in the present work,  but also the generation of new fractures in response to the evolving stress state from both thermo-and hydraulic perturbations. Future models might also include fracture roughness and solve the full equilibrium equations to estimate aperture changes that influence permeability. Recently, progress in this direction has been made using boundary element methods, multi-point stress approximations (MPSA) and the novel extended finite volume method (XFVM) \citep{norbeck2016,ucar2016,deb2016}. However, these models are not yet as computationally efficient as to allow an adaption for THERMAID. Currently, induced seismicity can not be quantified in terms of magnitude because slip on the fracture is not computed. Moreover, fracture slip can occur in a seismic or aseismic manner, thus further complicating the assessment of induced seismicity. These are all areas that we are currently pursuing in order to extend and refine THERMAID's capabilities. \\ 
Although this paper focuses on the application of enhanced geothermal systems, other possible applications for THERMAID include  seasonal thermal energy storage in fractured aquifers, and natural or anthropogenic fluid-driven earthquake sequences.  Furthermore, a wide range of research questions related to fractured reservoirs and their properties can be addressed using THERMAID. Even beyond the current model capabilities, we expect further applications and research opportunities because the open source code will allow a community to evolve and contribute to this common platform. The open source distribution and GNU GPL v3.0 license enables the scientific community to use and modify THERMAID to their needs. The implementation in \textit{MATLAB} ensures that even novice programmers can easily understand the underling equations and their implementation and develop their own numerical models based on the examples provided with THERMAID. Simulation of coupled processes in fractured reservoirs is becoming increasingly important in today's research. With THERMAID we present an alternative starting point from which new insight can be gained into the complex coupled processes in fractured domains in the subsurface.

\section*{Acknowledgements}
  We thank the Swiss National Fond (SNF), for the financial support through the grant ’NFP70:Energy Turnaround’ under the project no. 153971.
  
\section*{Computer Code Availability}
  \begin{itemize}
  	\item \textbf{Project name:}  THERMAID 
  	\item \textbf{Project home page:}  https://github.com/gujans/THERMAID
  	\item \textbf{Referenced archived version DOI:} 10.5281/zenodo.1175829
  	\item \textbf{Operating system(s):}  Platform independent
  	\item \textbf{Programming language:} MATLAB 
  	\item \textbf{Other requirements:} built and tested with MATLAB R2015b
  	\item \textbf{Licence:} GNU GPL v3.0 or later
  	\item \textbf{Any restrictions to use by non-academics:} The terms of the GNU GPL v3.0 or later apply.
  \end{itemize}
  
  \section*{Competing interests}
  The authors declare that they have no competing interests.
  
  \section*{Funding}
  This work was supported by the Swiss National Fond (SNF), for the financial support through the grant ’NFP70:Energy Turnaround’ under the project no. 153971.

\section*{Bibliography}

\bibliographystyle{abbrvnat}
\bibliography{Jansen_THERMAID.bib}      %

\end{document}